\documentclass[aps,prx,twocolumn,longbibliography,eqsecnum,floatfix,superscriptaddress]{revtex4-2}

\usepackage{amsmath}
\usepackage{amssymb}
\usepackage{mathtools}
\usepackage{amsfonts}
\usepackage{bm}

\usepackage{multibib}
\newcites{supp}{Supplementary References}
\usepackage{color}
\usepackage{xcolor}
\definecolor{blue}{rgb}{0.25, 0.3, 0.95}
\definecolor{red}{RGB}{0,100,0}
\definecolor{darkgreen}{rgb}{0.0, 0.5, 0.0}

\usepackage{braket}

\usepackage[caption=false]{subfig}

\newcommand{\phantomsubfloat}[1]{
    {
        \captionsetup[subfigure]{labelformat=empty}
        \subfloat[][]{#1}
    }%
}
\usepackage{graphicx} 

\newcommand{\avg}{\mathrm{avg}}
\newcommand{\typ}{\mathrm{typ}}
\newcommand{\toy}{\mathrm{toy}}
\newcommand{\cluster}{\mathrm{cluster}}
\newcommand{\lc}{\overline{\ell_{\cluster}}}

\usepackage{tikz}
\usepackage{textcomp}

\def\be{\begin{equation}}
\def\ee{\end{equation}}
\def\bea{\begin{eqnarray}}
\def\eea{\end{eqnarray}}

\usepackage[colorlinks,bookmarks=false,citecolor=red,linkcolor=blue,urlcolor=blue]{hyperref}

\begin{document}
\title{Breaking the chains: extreme value statistics and localization in random spin chains}
\author{Jeanne Colbois}
\email{jeanne.colbois@irsamc.ups-tlse.fr}
\affiliation{Laboratoire de Physique Th\'{e}orique, Universit\'{e} de Toulouse, CNRS, UPS, France}
\author{Nicolas Laflorencie}
\email{nicolas.laflorencie@irsamc.ups-tlse.fr}\affiliation{Laboratoire de Physique Th\'{e}orique, Universit\'{e} de Toulouse, CNRS, UPS, France}

\begin{abstract}
Despite a very good understanding of single-particle Anderson localization in one-dimensional (1D) disordered systems, many-body effects are still full of surprises, a famous example being the interaction-driven many-body localization (MBL) problem, about which much has been written, and perhaps the best is yet to come. Interestingly enough the non-interacting limit provides a natural playground to study non-trivial multiparticle physics, offering the possibility to test some general mechanisms with very large-scale exact diagonalization simulations. In this work, we first revisit the 1D many-body Anderson insulator through the lens of extreme value theory, focusing on the extreme polarizations of the equivalent spin chain model in a random magnetic field. A many-body-induced chain breaking mechanism is  explored numerically, and compared to an analytically solvable toy model. A unified description, from weak to large disorder strengths $W$ emerges, where the disorder-dependent average localization length $\xi(W)$ governs the extreme events leading to chain breaks. In particular, tails of the local magnetization distributions are controlled by $\xi(W)$. Remarkably, we also obtain a quantitative understanding of the full distribution of the extreme polarizations, which is given by a Fr\'echet-type law. In a second part, we explore finite interaction physics and the MBL question.
For the available system sizes, we numerically quantify the difference in the extreme value distributions between the interacting problem and the non-interacting Anderson case. Strikingly, we observe a sharp "extreme-statistics transition" as $W$ changes, which may coincide with the MBL transition.
\end{abstract}
\maketitle
\tableofcontents


\section{Introduction}
\label{sec:Intro}
\subsection{Context}
It is fairly well known that one-dimensional (1D) quantum systems are strongly influenced by the presence of random impurities or quenched disorder. 
This is firstly true at the single particle level where the simplest 1D nearest-neighbor hopping problem in a random potential displays the Anderson localization phenomenon whose hallmark is a real-space exponential decay of all eigenstates, regardless of the strength of the disorder~\cite{Mott_1961}. Another key property is the absence of transport at any energy in such localized systems, a direct consequence of destructive interferences of wave functions~\cite{Anderson_1958,50y}. Interestingly, despite an abundant corpus of well-established results in the non-interacting limit, it has proven to be notoriously difficult to translate them to the more realistic situation of finite interactions~\cite{fleishman_interactions_1980}, with the notable exception of zero-temperature physics thanks to bosonization and RG~\cite{Giamarchi_1988}. For the more general case of finite temperature, in principle involving all many-body excitations, the celebrated many-body localization (MBL) problem has attracted an enormous interest over the past  decades~\cite{altshuler_quasiparticle_1997,jacquod_emergence_1997,gornyi_interacting_2005,basko_metalinsulator_2006,znidaric_many-body_2008,pal_many-body_2010,bardarson_unbounded_2012,luitz_many-body_2015,schreiber_observation_2015,imbrie_many-body_2016,smith_many-body_2016,choi_exploring_2016,roushan_spectroscopic_2017, alet_many-body_2018,abanin_many-body_2019}. Remarkably, a great deal of effort has been put into numerical simulations of interacting 1D models, mostly the random-field Heisenberg spin chain~\cite{pal_many-body_2010,luitz_many-body_2015}, for which an ergodicity-breaking transition is expected at strong enough disorder~\cite{luitz_many-body_2015,doggen_many-body_2018,chanda_time_2020,sierant_thouless_2020,abanin_distinguishing_2021}.

Nevertheless, the objective of this work is not only oriented towards interacting MBL. Instead we propose to make a first detour to the non-interacting many-body Anderson insulator through the less visited question of extreme value theory. We will concentrate on the effect of a random magnetic field on extreme polarizations in one of the simplest disordered spin chain models, namely the spin-$\frac{1}{2}$ random-field XX chain, governed for $L$ sites on a ring by
\begin{equation}
\mathcal{H}_{\rm xx} = \sum_{i=1}^{L} J\left(S_i^x S_{i+1}^x + S_i^y S_{i+1}^y \right) -h_i S_i^{z},
\label{eq:XX}
\end{equation}
which is the non-interacting limit of the 1D Heisenberg model.
Our goal is to develop a theoretical understanding of the extreme magnetizations at any temperature, first for this non-interacting case, building on a combination of large-scale exact diagonalization with an extreme value theory (EVT) analysis of local observables. 
Next, we aim to extend our conclusions to the finite interaction problem, thus offering us the possibility to strengthen our understanding of MBL physics.

Univariate EVT~\cite{Gumbel_1958,leadbetter_extremes_1983,coles_introduction_2001,DeHaan_2006, Majumdar_2020} has proven useful in fields as diverse as climate science~\cite{wigley_climatology_1985,wigley_effect_2009,cooley_extreme_2009,naveau_statistical_2005} and extreme environmental events~\cite{Tippett_2016, tabari_extreme_2021}; structural risks and related safety measures~\cite{Osti_1994,Makkonen_2008}; athletic records~\cite{Gembris_2002,Gembris_2007,Einmahl_2011, Spearing_2021}; finance~\cite{Novak_2012, gkillas_application_2018}; and statistical physics~\cite{fortin2015, Majumdar_2020,bramwell_1998}. In particular, disordered condensed matter systems~\cite{Biroli_2007} offer a very rich playground for EVT of correlated or uncorrelated random variables, with notable examples such as random matrix theory~\cite{tracy_orthogonal_1996,fydorov_freezing_2014}, height distribution of surfaces~\cite{fortin2015, majumdar_exact_2004,dotsenko_bethe_2010,calabrese_free-energy_2010} and spin or structural glasses~\cite{bouchaud_universality_1997, derrida_random-energy_1981,derrida_generalization_1985,derrida_solution_1986, Castellana_2014}.
Disordered spin chains represent a potentially very interesting field of application for EVT, with the  notable example of the random transverse-field Ising chain model~\cite{fisher_critical_1995} for which EVT provides very sharp predictions for the distribution of the lowest finite-size energy gap~\cite{fisher1998,Juhasz_2006,Kovacs_2021,Kao_2022}. EVT can also be used to study localization-delocalization transitions: in the (single-particle) Anderson problem with long-range couplings, extreme values of the eigenfunctions capture the localized or delocalized nature of the wavefunctions~\cite{Falcao_2022}. Other recent examples of studies include the ground-state energy distribution of quantum disordered chains~\cite{Buijsman2022}, the extremal statistics of entanglement spectra across the 1D MBL transition~\cite{Buijsman_2019,Samanta2020}, the superfluid fraction in a one-dimensional Bose gas~\cite{Albert_2020} or the probability distribution of the commutators of almost-conserved local operators~\cite{pancotti_almost_2018}.

\begin{figure*}[tp]
    \centering
    \includegraphics[width=1.45\columnwidth]{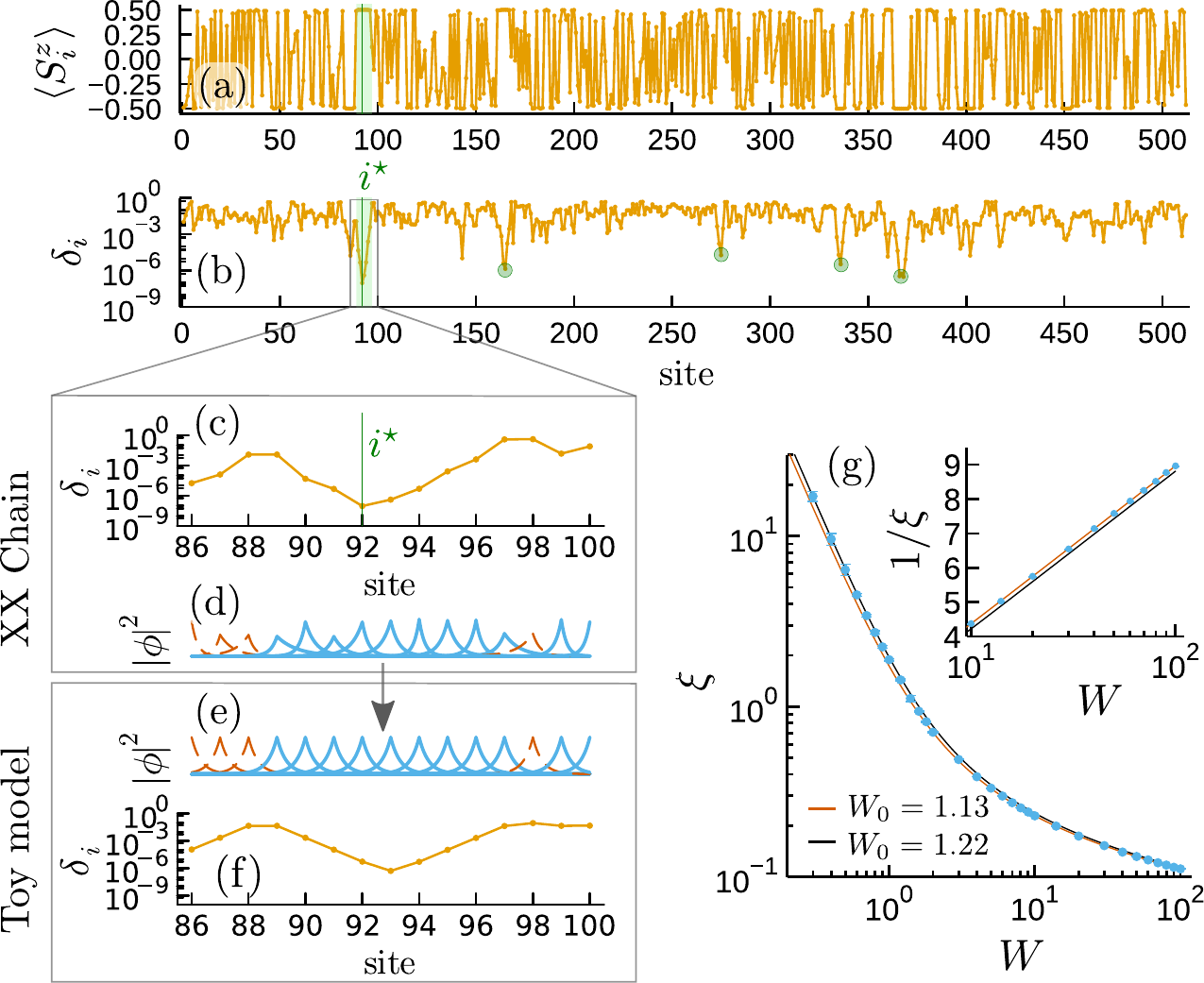}
	\phantomsubfloat{\label{fig:toy1_a}}
	\phantomsubfloat{\label{fig:toy1_b}}
	\phantomsubfloat{\label{fig:toy1_c}}
	\phantomsubfloat{\label{fig:toy1_d}}
	\phantomsubfloat{\label{fig:toy1_e}}
	\phantomsubfloat{\label{fig:toy1_f}}
	\phantomsubfloat{\label{fig:toy1_g}}
        \caption{Microscopic mechanism of the chain breaking in the non-interacting case, Eq.~\eqref{eq:XX}, illustrated for a single $L=512$ sites  sample with a disorder strength $W=5$. \protect\subref*{fig:toy1_a} Expectation value of the local magnetization $\langle S_i^z\rangle$ along the chain, computed with ED for an eigenstate in the middle of the spectrum:  we observe seemingly random oscillations between $ \pm 1/2$. \protect\subref*{fig:toy1_b} Same as panel \protect\subref*{fig:toy1_a} but for the deviations  with respect to perfect polarization $\delta_i=1/2-|\langle S_i^z\rangle|$, plotted in log-scale. Green circles highlight the sites with the smallest deviations (strongly polarized spins). The most polarized site $i^{\star}=92$ is indicated by a vertical green line and the cluster containing it by a green region in panels~\protect\subref*{fig:toy1_a} and~\protect\subref*{fig:toy1_b}. \protect\subref*{fig:toy1_c} Zoom over the region surrounding $i^\star$: one clearly sees a short-range correlation of the $\delta_i$'s in its vicinity. \protect\subref*{fig:toy1_d} Microscopic mechanism at the origin of the chain breaking: the most polarized site lies in a series of $\ell_{\rm max}\approx \frac{\ln L}{\ln 2}=9$ consecutively occupied orbitals $\phi_m$, represented by full blue lines, orange dashed lines representing the unoccupied orbitals. Panel \protect\subref*{fig:toy1_d} shows exponential fits to the exact single-particle states, not necessarily symmetric, while panel \protect\subref*{fig:toy1_e} represents the corresponding toy model description with all $\phi_m$ in Eq.~\eqref{eq:phim} having the same localization length $\xi$; the resulting deviations are shown in panel \protect\subref*{fig:toy1_f}. \protect\subref*{fig:toy1_g} Disorder dependence of the localization length $\xi(W)$ computed from the Lyapunov exponent (see~\cite{SM} Sec.~S2) averaged over the density of states. The continuous lines correspond to the analytical ansatz $\xi^{-1}=\ln[1 + ({W}/{W_0})^2]$, with $W_0=1.13, 1.22$ (see text Eq.~\eqref{eq:XiFit} and below). The inset shows the same data for $1/\xi$ at large disorder. \label{fig:toy1}}
\end{figure*}

In this MBL context, the full distributions have been studied for the entanglement entropy~\cite{Bauer_2013,Luitz_2016,Yu_2016,PhysRevX.7.021013}, while much less authors have addressed the pairwise correlations at short~\cite{Colmenarez_19} or long distance~\cite{pal_many-body_2010}. Also much less studied~\cite{Laflorencie_2020,Hopjan_21} are the statistics of one of the simplest local observables: the local magnetization $\langle S_i^z\rangle$. Although simple, this quantity can be of deep experimental relevance, since inhomogeneous magnetic moment profiles can be resolved directly by NMR spectroscopy~\cite{Tedoldi_1999,Kodama_2002,Kramer_2013,Zhou_2017}. On the theory side, only a few  works have addressed magnetization statistics across the ergodic-MBL transition in disordered spin chains, mostly at a qualitative level, for instance observing a U-shape distribution in the localized regime, contrasting with a Gaussian law in the ergodic phase~\cite{Khemani_2016,Lim_2016,Dupont_2019a,Hopjan_2020}. Interestingly, Ref.~\onlinecite{Lim_2016} noticed a power-law decay of the probability to observe vanishing local magnetization $P(0)\sim 1/W$, where $W$ controls the disorder strength, while at the same time tail singularities develop with $W$ for $\langle S_i^z\rangle\approx \pm\frac{1}{2}$. Motivated by these results, we propose to use extreme value statistics to obtain a more quantitative description of this crossover. 
\subsection{Chain breakings in a nutshell: toy model as a warmup}
\label{sec:CB}
For a finite chain governed for instance by the quantum XX model Eq.~\eqref{eq:XX}, the local spin densities $\langle S_i^z\rangle$ can get very close to $\pm \frac{1}{2}$ upon increasing the strength of the random field $W$, but never reach perfect (classical) polarization. This is an obvious consequence of the quantum fluctuation terms in the Hamiltonian. However, such a classical spin freezing may happen when the thermodynamic limit is taken, leading to a chain break, i.e. some weak links effectively flowing to zero strength when $L\to \infty$. This was first observed and discussed in Ref.~\onlinecite{Laflorencie_2020} when investigating the extreme statistics in random Heisenberg chains, where it was concluded that quantum localization is associated to a chain breaking mechanism. Formally speaking, this means a classical freezing at the thermodynamic limit, i.e. the most polarized site $i^\star$ will be  such that
\be
\langle S_{i^\star}^z\rangle \underset{L\to \infty}{\longrightarrow} \pm\frac{1}{2}.
\ee
Let us  first give a heuristic explanation for this freezing phenomenon in the non-interacting limit Eq.~\eqref{eq:XX}, which describes free fermions in a random potential. In the presence of disorder, all single-particle fermionic eigenstates $\phi_m$ are Anderson localized. Following Refs.~\onlinecite{Dupont_2019, Laflorencie_2020}, we model these eigenstates by a simple exponential 
\be
|\phi_m(i)|^2\sim \exp\left(-\frac{|i-i_0^m|}{\xi_m}\right)
\label{eq:phim}
\ee
for all orbitals $m$, with $\xi_m$ and $i_0^m$ the corresponding localization lengths and centers.  For a given filling fraction $0<\nu<1$, the real-space  density at a site $i$ is given by 
\be
\langle n_i\rangle =\sum_{m_{\rm oc.}}|\phi_{m_{\rm oc.}}(i)|^2,
\ee
where the sum is performed over the $\nu L$ occupied fermionic levels $m_{\rm oc.}$. In our toy model description, we then make the defining assumption that all  orbitals have the same localization length $\xi_m\equiv\xi$. This strong assumption, whose validity will be discussed  in Sec.~\ref{sec:ToyModel} in the light of systematic ED results on the XX chain, allows us to easily extract simple predictions. For instance, as a direct consequence, the maximal (resp. minimal) fermionic density is expected to occur in the middle of the longest region of $\ell_{\max}$ consecutive sites that are occupied (unoccupied) by an orbital~\footnote{A site occupied by an orbital means that the localization center $i_0^m$ is precisely located at this site.}. At half-filling $\nu=1/2$, a configuration with $\ell$ consecutive occupied (or empty) sites occurs with a probability proportional to $ 2^{-\ell}$, which, for a finite chain of length $L\gg 1$ yields $\ell_{\max}\approx \ln L/\ln 2$. 
Back to the spin language, the minimal deviation from perfect polarization $\delta_{\min}\equiv \frac{1}{2}-\left|\langle S_{i^\star}^z\rangle\right|$, is then given by
\be
\delta_{\min}(L)\sim \exp\left(-\frac{\ell_{\max}}{2\xi}\right)\sim L^{-\frac{1}{2 \xi \ln 2}},
\label{eq:deltamin_lmax}
\ee
which  defines the disorder-dependent freezing exponent
\be
    \gamma=\frac{1}{2\xi \ln 2}.
    \label{eq:gamma_xi}
\ee
This simple reasoning, illustrated in Fig.~\ref{fig:toy1} \protect\subref*{fig:toy1_a} to~\protect\subref*{fig:toy1_f}, will be further discussed below in the paper, together with large-scale numerical simulation results.
\subsection{Summary of the results and outline of the paper}
In the rest of the paper we will explore various aspects of extreme polarizations in random-field spin chains at infinite temperature. First, in Sec.~\ref{sec:ED} 
 we use exact diagonalization (ED) to investigate large chains in the non-interacting limit for which the statistics of local magnetizations, and particularly of its extreme values are probed in great details. Finite-size scaling of the spin freezing process will be systematically studied for a broad range of disorder strengths, and compared to the concomitant weak link formation leading to chain breaks in the thermodynamic limit. The associated disorder-dependent spin freezing and weak-link exponents will be analyzed, and  compared to the simple the toy model expectation Eq.~\eqref{eq:gamma_xi}. 
A key result, presented in Sec.~\ref{sec:Frechet_main}, is that throughout the entire Anderson insulating regime, the  distributions of  extreme polarizations fall in the Fr\'echet subclass of  generalized extreme value theory.
 
 The toy model is then further explored in Sec.~\ref{sec:ToyModel} for the  many-body Anderson insulator, with a particular focus on its capability to provide quantitative results against ED data. The minimal deviations (from perfect polarization) are nicely described  by a scaling-law inferred from the toy model, with a non-trivial scaling variable ${\overline{\ell_{\rm cluster}}}/\xi$, which is the ratio between the length of the cluster hosting the most polarized spin ${\overline{\ell_{\rm cluster}}}$ and the average localization length $\xi$. Interestingly, the same toy model also provides quantitative results to explain several features in the full distribution functions, as we discuss in Sec.~\ref{sec:TM_Dist}.

 Finally, Sec.~\ref{sec:MBL} addresses the MBL question through the lens of extreme value theory, building on both shift-invert ED data and an interacting toy model description from which we first make analytical conjectures valid at strong disorder (Sec.~\ref{sec:ITM}). We then compare numerical data between XX and Heisenberg Hamiltonians for both the typical extreme deviations (Sec.~\ref{sec:NUMMBL}) and their full probability distribution functions (Sec.~\ref{sec:EVSMBL}). The Kullbach-Leibler divergence, used to quantify the differences / similarities between the interacting and non-interacting distributions, exhibits a sharp change for a critical disorder regime $W\sim 4-7$, thus suggesting an extreme-statistics transition, which might coincide with the MBL transition.

We summarize and conclude in Sec.~\ref{sec:Discussion}. 
More details on some technical aspects of the calculations are provided as Supplemental Material~\cite{[{See Supplemental Material at }]SM}.

\section{Exact diagonalization of the many-body Anderson insulator}
\label{sec:ED}
\subsection{The non-interacting XX model}
\label{sec:NIModel}
We start with the random-field XX chain model Eq.~\eqref{eq:XX}, which can be recast as a tight-binding chain of spinless fermions with random on-sites energies, governed (up to a constant) by 
\be
    \mathcal{H}_{\rm xx}=\sum_{i=1}^{L} \frac{J}{2} \left(c_i^{\dagger} c_{i+1}^{\vphantom{\dagger}} + c_{i+1}^{\dagger}c_{i}^{\vphantom{\dagger}}\right) -h_i n_i.
    \label{eq:HFermions}
\ee
Throughout the paper, we consider $J=1$ and a uniform box distribution for the random fields $h_i$, i.e.
\begin{equation}
    \mathcal{P}(h) = \begin{cases}
    \frac{1}{2W} & \text{ if } h \in [-W, W],\\
    0 & \text{ otherwise},
    \end{cases}
\end{equation}
which has a finite variance $W^2/3$. The model has a $U(1)$ symmetry corresponding to the total magnetization (particle number) being conserved. Unless otherwise stated, we consider the spin-1/2 chain in the $S^{z}_{\mathrm{tot}} = 0$ sector (half-filling in the fermion language). Exact diagonalization of the above quadratic Hamiltonian are performed over several thousands (typically $10^4$) independent samples for periodic chains of lengths spanning a few orders of magnitudes (typically from $L=8$ to $L=4096$). For each disordered sample, new fermionic operators are numerically obtained from standard ED: $b_m=\sum_{i=1}^{L}\phi_m(i)c_i$, such that the free-fermion Hamiltonian takes a diagonal form
\be
{\cal{H}}=\sum_{m=1}^{L}{\cal{E}}_m b_m^{\dagger} b_{m}^{\vphantom{\dagger}}.
\label{eq:diag}
\ee
All single-particle orbitals $\phi_m(i)$ are spatially localized for any finite disorder strength $W>0$.
We focus on the infinite temperature limit, corresponding to the middle of the many-body spectrum with $\epsilon = 0.5$ (see ~\cite{SM} Sec.~S1 for some more details).

\subsection{Distribution of local magnetizations}
\label{sec:Mdistrib}
\begin{figure}[t!]
    \centering
    \includegraphics[clip=true,width=\columnwidth]{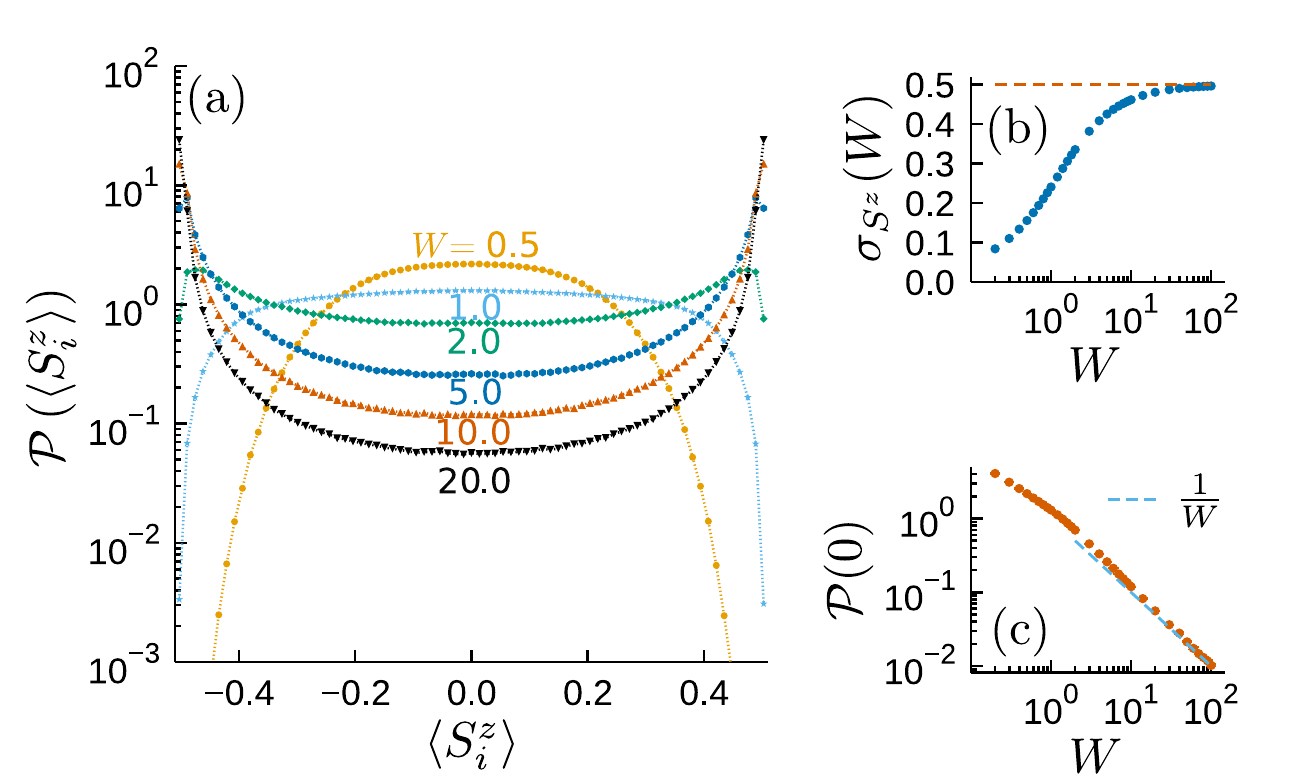}
	\phantomsubfloat{\label{fig:Occupations_a}}
	\phantomsubfloat{\label{fig:Occupations_b}}
	\phantomsubfloat{\label{fig:Occupations_c}}
    \caption{ \protect\subref*{fig:Occupations_a}~Histograms of the local magnetizations $\langle S_i^{z} \rangle$ at half-filling and infinite temperature, shown for various disorder strengths. ED data for $L=512$, periodic boundary conditions, and $10^4$ random-field realizations. ${\cal{P}}$ gradually changes from an arch to a U-shaped curve with singular tails upon increasing $W$.  \protect\subref*{fig:Occupations_b}~The standard deviation of the distribution of magnetizations as a function of the disorder strength (symbols) saturates at $1/2$ for strong disorder. \protect\subref*{fig:Occupations_c}~The weight of the distribution at zero magnetization decreases with increasing disorder (points). The dashed line represents the expected zero magnetization weight in the strong disorder limit $\sim 1/W$. }
    \label{fig:Occupations}
\end{figure}

We first look at the distribution of the on-site magnetizations $\langle S_j^z \rangle$, equivalent to the fermionic occupations $\langle n_j \rangle = \langle S_j^{z} \rangle + 1/2$. In Fig.~\ref{fig:Occupations_a} we observe the development of singular tails for the extremal values $\langle S_i^z \rangle\to\pm 1/2$ upon increasing the disorder strength $W$, with distributions becoming bimodal in the large disorder limit. The standard deviation $\sigma_{S^z}(W)$, plotted in Fig.~\ref{fig:Occupations_b}, grows continuously with $W$ and tends to saturation at the bimodal distribution plateau $\sigma_p\to 1/2$ for strong disorder. At the same time, the weight at zero magnetization vanishes ${\cal{P}}(0)\sim 1/W$, see Fig.~\ref{fig:Occupations_c}, similarly to the case with interactions~\cite{Lim_2016,Dupont_2019a,Dupont_2019,Laflorencie_2020}.

\begin{figure}[h!]
    \centering
    \includegraphics[width=0.85\columnwidth]{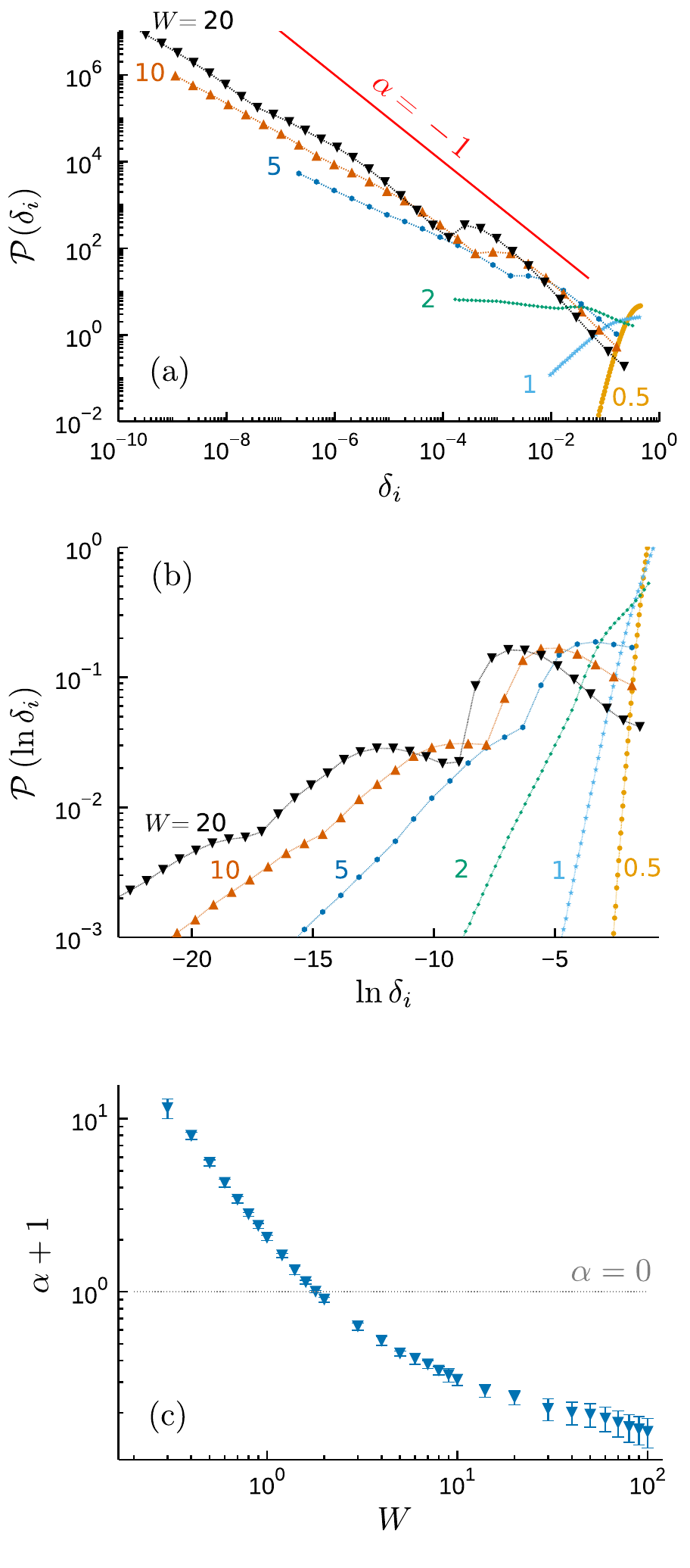}
	\phantomsubfloat{\label{fig:LnDeviationsHistogram_a}}
	\phantomsubfloat{\label{fig:LnDeviationsHistogram_b}}
	\phantomsubfloat{\label{fig:alpha}}
    \caption{Histogram of the deviations $\delta_i$ to perfect polarization Eq.~\eqref{eq:DeltaDef}, for various disorder strengths $W$. Same ED data as shown in Fig.~\ref{fig:Occupations_a}. \protect\subref*{fig:LnDeviationsHistogram_a}~The probability of small deviations exhibit clear power-law scaling in log-log scale, Eq.~\eqref{eq:PowerLaw}. The line indicates the most singular situation $\alpha = -1$. \protect\subref*{fig:LnDeviationsHistogram_b}~Distribution of the logarithm of the deviations, showing the corresponding exponential scaling. Note the striking structures  for strong disorder away from the small-$\delta$ tails regime, structures which are invisible in the linear scale of Fig.~\ref{fig:Occupations_a}. \protect\subref*{fig:alpha}~Disorder dependence of the power-law exponent $\alpha$ Eq.~\eqref{eq:PowerLaw}, obtained from fitting the tails of the distributions of $\ln\delta_i$ (see also Fig.~\ref{fig:Exponents}). $\alpha$ switches sign at $W^*\approx 1.8$.}
    \label{fig:LnDeviationsHistogram}
\end{figure}

In contrast, for weak disorder the distribution ${\cal{P}}(\langle S_j^z \rangle)$ changes from the strong disorder U-shaped convex form and instead becomes concave.
Therefore, the probability of observing strongly polarized sites having $\langle S_i^z \rangle$  close to $\pm 1/2$ exhibits seemingly very different behaviors as a function of the disorder strength $W$.  At strong disorder it has previously been argued~\cite{Dupont_2019,Laflorencie_2020} that the distribution diverges as a power-law $\sim \delta^{-|\alpha|}$, where $\delta$ is the deviation from perfect polarization, defined at each site 
by 
\be
\delta_i = \frac{1}{2}-|\langle S_i^z \rangle|.
\label{eq:DeltaDef}
\ee
Interestingly, we numerically find that the algebraic tail at $\delta\to 0$ remains over the entire regime of disorder strengths, as shown in Fig.~\ref{fig:LnDeviationsHistogram} where the form
\begin{equation}
    \label{eq:PowerLaw}
    {\cal{P}}(\delta) \sim \delta^\alpha \quad \Leftrightarrow     \quad {\cal{P}}(\ln \delta) \sim \exp\left[{(1+\alpha) \ln\delta}\right],
\end{equation}
is always observed. The $W$-dependent exponent $\alpha\in ]-1,\infty[$ is found to continuously vary with the disorder strength, see Fig.~\ref{fig:alpha} where we observe a saturation  $\alpha\to -1$ in the large $W$ regime, while $\alpha\to \infty$ at weak disorder.
Although a qualitative change seems to occur when the exponent $\alpha$ switches sign at $W^*\approx 1.8$, at least with respect to the divergent or vanishing character of ${\cal{P}}(\delta\to 0)$, the physical regime is however the same on the two sides of $W^*$ which are both described by the same Anderson localization phenomenology.

\subsection{Extreme value theory and chain breaks}
Power-law tails in the distributions yield strong consequences for the \emph{minimal} deviation $\delta_{\min} = \min_{j} \delta_j$,  corresponding to the most polarized site. Indeed, following the Gnedenko's classical law of extremes, it is well-known~\cite{Gnedenko_1943,Majumdar_2020} that for such an algebraic distribution of {\it{independent}} random variables, the statistics of the extremes will fall into the Fr\'echet distribution universality class. This will be further developed below, in Sec.~\ref{sec:Frechet_main} (see also Supplemental Material~\cite{SM}, Sec.~S3). Before this, we  first focus on the finite-size scaling properties of the extreme polarizations for finite chains of length $L$. In the following we will further assume weak correlations between the $\delta_i$ random variables, which is well-justified as we discuss in more detail in Sec.~\ref{sec:corr}.
\begin{figure}
    \centering	
    \includegraphics[width=0.93\columnwidth]{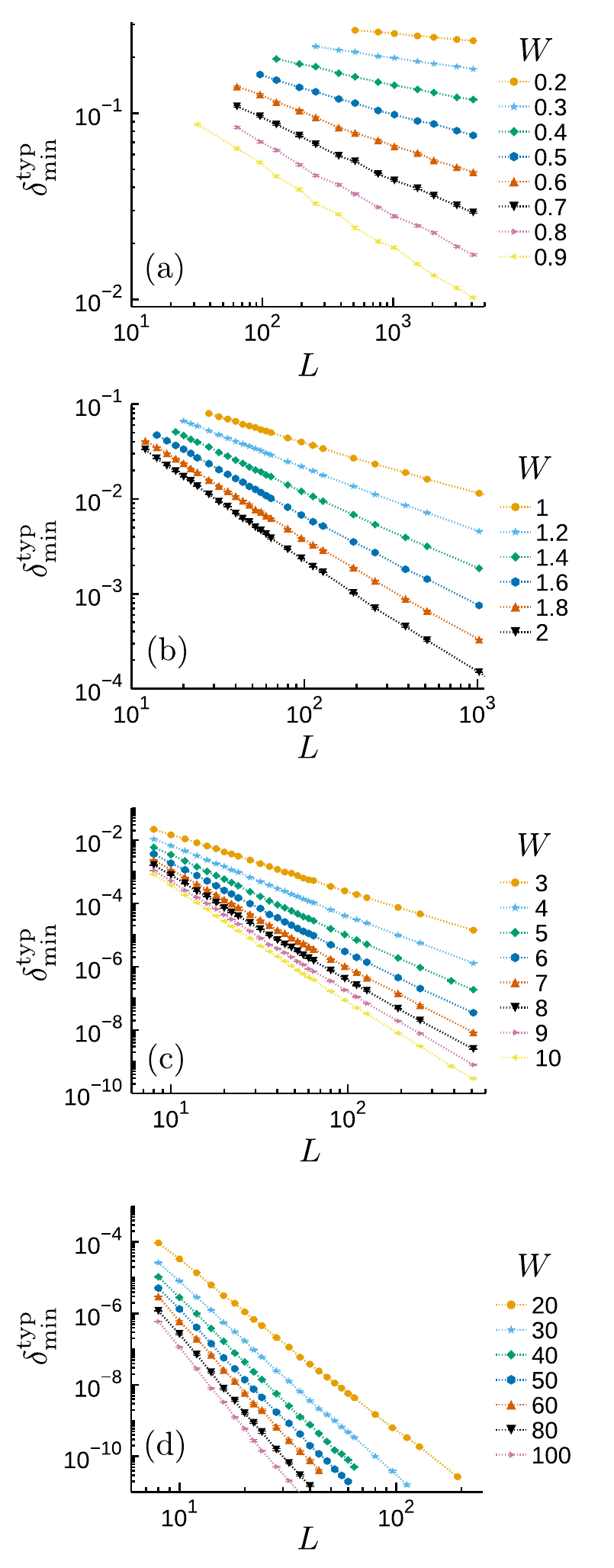}
	\phantomsubfloat{\label{fig:SizeDepDeltaMin_a}}
	\phantomsubfloat{\label{fig:SizeDepDeltaMin_b}}
	\phantomsubfloat{\label{fig:SizeDepDeltaMin_c}}
	\phantomsubfloat{\label{fig:SizeDepDeltaMin_d}}
    \caption{Infinite temperature ED data for the typical value of the minimal deviations, $\exp[\overline{\ln\delta_{\min}}]$ as a function of the system size $L$ and various disorder strengths $W$, plotted for $L \gg \xi(W)$. $10^3$ field realizations were used for panel~\protect\subref*{fig:SizeDepDeltaMin_a} and $10^4$ field realizations for panels~\protect\subref*{fig:SizeDepDeltaMin_b} to~\protect\subref*{fig:SizeDepDeltaMin_d}. Dotted lines are a guide to the eye. The extracted freezing exponent $\gamma_\typ$ Eq.~\eqref{eq:gammatyp} is shown in Fig.~\ref{fig:Exponents}.}
    \label{fig:SizeDepDeltaMin}
\end{figure}

\subsubsection{Spin freezing}
\label{sec:SpinFreezing}
For a finite system of size $L$, the typical behavior of the minimal deviation from perfect polarization $\delta_{\min}(L)$ can be obtained from the simple argument that it must occur exactly once in the chain, such that
\begin{equation}
\label{eq:PowLawMin}    
    \int_0^{\delta_{\rm{min}}(L)}\mathcal{P}(\delta) \, \mathrm{d} \delta \sim \frac{1}{L}.
\end{equation}
Therefore, the algebraic distribution Eq.~\eqref{eq:PowerLaw} yields the following finite-size decay (which we will recover from a more systematic EVT analysis in Sec.~\ref{sec:Frechet_main})
\be
\delta_{\rm{min}}(L) \sim L^{-\frac{1}{1+\alpha}},
\label{eq:deltamin}
\ee
thus linking the exponent $\alpha$ of the distribution with the classical freezing exponent 
\be
\gamma=\frac{1}{1+\alpha}.
\label{eq:gamma_alpha}
\ee
This implies that  for any $\alpha > -1$, the minimal deviation is expected to vanish in the thermodynamic limit $L \rightarrow \infty$, with a freezing exponent $\gamma>0$, and the most polarized site $i^\star$ eventually becomes classically frozen $\langle S_{i^\star}^{z}\rangle=\pm 1/2$, in the sense that the quantum fluctuations have completely disappeared.

This algebraic decay for $\delta_{\min}(L)$ Eq.~\eqref{eq:deltamin} is checked numerically by collecting the most polarized site for $10^4$ samples and various sizes (we only use $10^3$ samples at weak disorder where sample-to-sample fluctuations are weaker, thus allowing to reach larger sizes). To account for the possible absence of self-averaging of these quantities, we consider two definitions of the freezing exponent $\gamma$ related to the typical and the average value of the minimal deviation:
\bea
	\label{eq:gammatyp}
	\delta_{\min}^{\typ} &=& e^{\overline{\ln(\delta_{\min})}} \sim L^{-\gamma_{\typ}}\\
	\delta_{\min}^{\avg} &=& \overline{\delta_{\min}} \sim L^{-\gamma_{\avg}}.
	\label{eq:gammaavg}
\eea
In Fig.~\ref{fig:SizeDepDeltaMin}, we show ED results for the typical value of the minimal deviations, which exhibit very clear power-law decays with $L$. This remarkably remains true for a very broad range of disorder strengths $W$, as illustrated by the different panels of Fig.~\ref{fig:SizeDepDeltaMin}. 
For weak disorder, $W<1$ [Fig.~\ref{fig:SizeDepDeltaMin_a}], the asymptotic regime is only reached at sufficiently large scale: typically we expect this scaling regime when $L\gg \xi$, with $\xi$ the localization length averaged over  orbitals and  disorder [Fig.~\ref{fig:toy1_g}]. Even at larger disorders, a careful look shows that the exponent of the power-law decay changes slightly with the increasing system size. This effect is at least partially due to the appearance of bumps in the distribution of $\delta_{\min}$ for strong disorder and modest system sizes (\cite{SM} Sec.~S3), and is more pronounced in the decay of the average minimal deviation. This is taken into account in the evaluation of the freezing exponent $\gamma_{\typ, \avg}$ by fitting ranges of six sizes (at $W<1$ and $W>30$) and eight sizes ($1\leq W \leq 30$) every two sizes, and using the four last fits (largest sizes) to determine the exponent, and all fits to determine the error on the exponent (see Fig.~\ref{fig:Exponents}). The extracted freezing exponent $\gamma_\typ$ clearly increases with $W$, as we will further discuss below in Sec.~\ref{sec:Exponents}. Before this we briefly address the chain breaking mechanism by looking at the spin-spin correlations  in the vicinity of the most polarized site.

\subsubsection{Weak links and chain breaks}
The power-law decay of the minimal deviation suggests that in the thermodynamic limit, the chain will eventually be cut upon increasing the system size. This behavior may remind us of single-impurity Kane-Fisher physics~\cite{Kane_92,Eggert_92,Lemarie_2019}, where the system flows towards an open chain fixed point. Another way to envision these weak links is to see them as entanglement bottlenecks~\cite{Luitz_2016}, nearly frozen spins being disentangled from the rest of the system. It is well-known that upon increasing disorder the distributions of entanglement entropies display a growing peak at zero entropy (with or without interaction) ~\cite{Luitz_2016,Yu_2016,Laflorencie_2022}, which is directly related to the singularity ${\cal{P}}(\delta\to0)$~\cite{Laflorencie_2020}.

\begin{figure}[h!]
    \centering
    \includegraphics[width=\columnwidth]{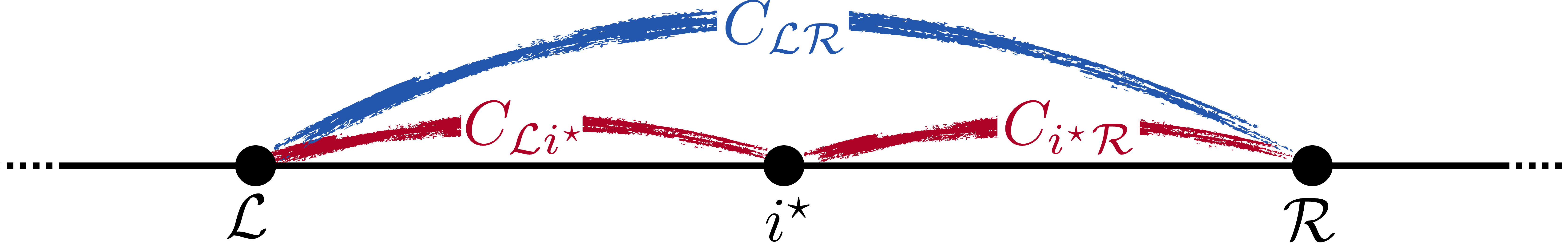}
    \caption{Schematic illustration of the pairwise spin correlations at short distance in the close vicinity of the most polarized site $i^\star$, surrounded by $\cal L$ (left) and $\cal R$ (right) sites.}
    \label{fig:Cistar}
\end{figure}

An explicit way to investigate such entanglement bottlenecks, or weak links, is to look at the gradual breaking of the chain when $L$ is increased. To do so, we compute (see~\cite{SM} Sec.~S1 for some details) the connected pairwise spin correlations in the close vicinity of the most polarized site $i^{\star}$, as illustrated in Fig.~\ref{fig:Cistar}.

\begin{figure}
    \centering	
    \includegraphics[width=.9\columnwidth]{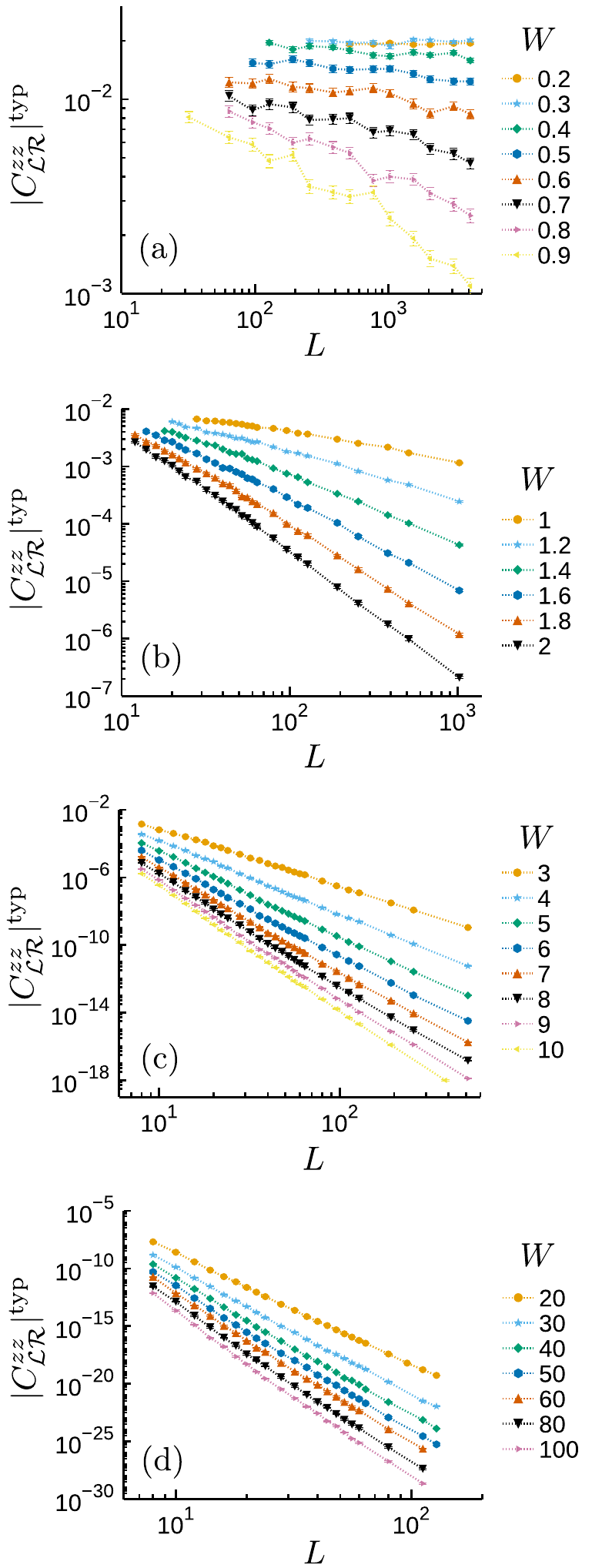}
	\phantomsubfloat{\label{fig:SizeDepCzz_a}}
	\phantomsubfloat{\label{fig:SizeDepCzz_b}}
	\phantomsubfloat{\label{fig:SizeDepCzz_c}}
	\phantomsubfloat{\label{fig:SizeDepCzz_d}}
    \caption{Power-law decay of the typical value of the connected longitudinal correlations : $\left|C^{zz}_{\cal LR}\right|$ measured across $i^{\star}$ the most polarized site, as illustrated in Fig.~\ref{fig:Cistar}. Infinite temperature ED data plotted as a function of the system size $L$ for various disorder strength $W$. Dotted lines are a guide to the eye. The extracted weak-link exponent $\theta_{\typ}$ is shown in Fig.~\ref{fig:Exponents} (except for $W < 0.5$ where it becomes too difficult to determine).}
    \label{fig:CorrelationsDecay}
\end{figure}

We numerically check that the longitudinal components
\be
C^{zz}_{ij}= \langle S^z_{i}S^z_{j} \rangle -\langle S^z_{i}\rangle\langle S^z_{j} \rangle
\ee
display a similar algebraic decay with $L$ for the three cases $(i\,j)=({\cal{L}}\,i^\star)$, $(i^\star\,{\cal{R}})$, as well as $({\cal{L}}\,{\cal{R}})$ which is shown in Fig.~\ref{fig:CorrelationsDecay} for the same parameters as the ones of Fig.~\ref{fig:SizeDepDeltaMin}. The typical weak-link correlation decays as
\begin{equation}
\label{eq:DefMu}
\left|C^{zz}_{\cal LR}\right|^{\typ} \sim L^{-\theta_{\typ}},
\end{equation}
which defines the weak-link exponent $\theta$ \footnote{We do not show here the exponent $\theta_{\avg}(W)$ corresponding to the decay of the \emph{average} of these spin-spin correlations. Indeed, as we discuss in~\cite{SM} Sec.~S1, due to rare resonances across the most polarized site contributing $|C^{zz}_{\cal LR}| \approx 0.25$, there is a slow decay of the average correlations with $L$.}.

It is straightforward to connect the above behavior with the spin freezing of Eq.~\eqref{eq:gammaavg}.  Indeed, one can model two nearly up-polarized sites using  the approximate quantum sate  
\begin{equation}
	\ket{\Psi}  \propto \ket{\uparrow_{i} \uparrow_{j}} 
	 + \epsilon_i \ket{\downarrow_{i} \uparrow_{j}}
	 + \epsilon_j \ket{\uparrow_{i} \downarrow_{j}},
\end{equation}
with $\epsilon_i\sim \epsilon_j\ll 1$. For such an ansatz wave-function, a simple calculation yields for the deviations $\delta_p\approx \epsilon_p^2\sim L^{-\gamma}$, and a weak-link correlation
\be
	\left|C^{zz}_{ij}\right|\approx\left(\epsilon_i\epsilon_j\right)^2\sim L^{-2\gamma},\ee
which leads to the following relation between the exponents 
\be
\label{eq:gamma_theta}
\theta = 2\gamma,
\ee
expected to describe the strong disorder regime, as we nicely check below in  Fig.~\ref{fig:Exponents}.

\subsubsection{Freezing exponents}
\label{sec:Exponents}

\begin{figure}[bp]
    \centering
    \includegraphics[width=.925\columnwidth]{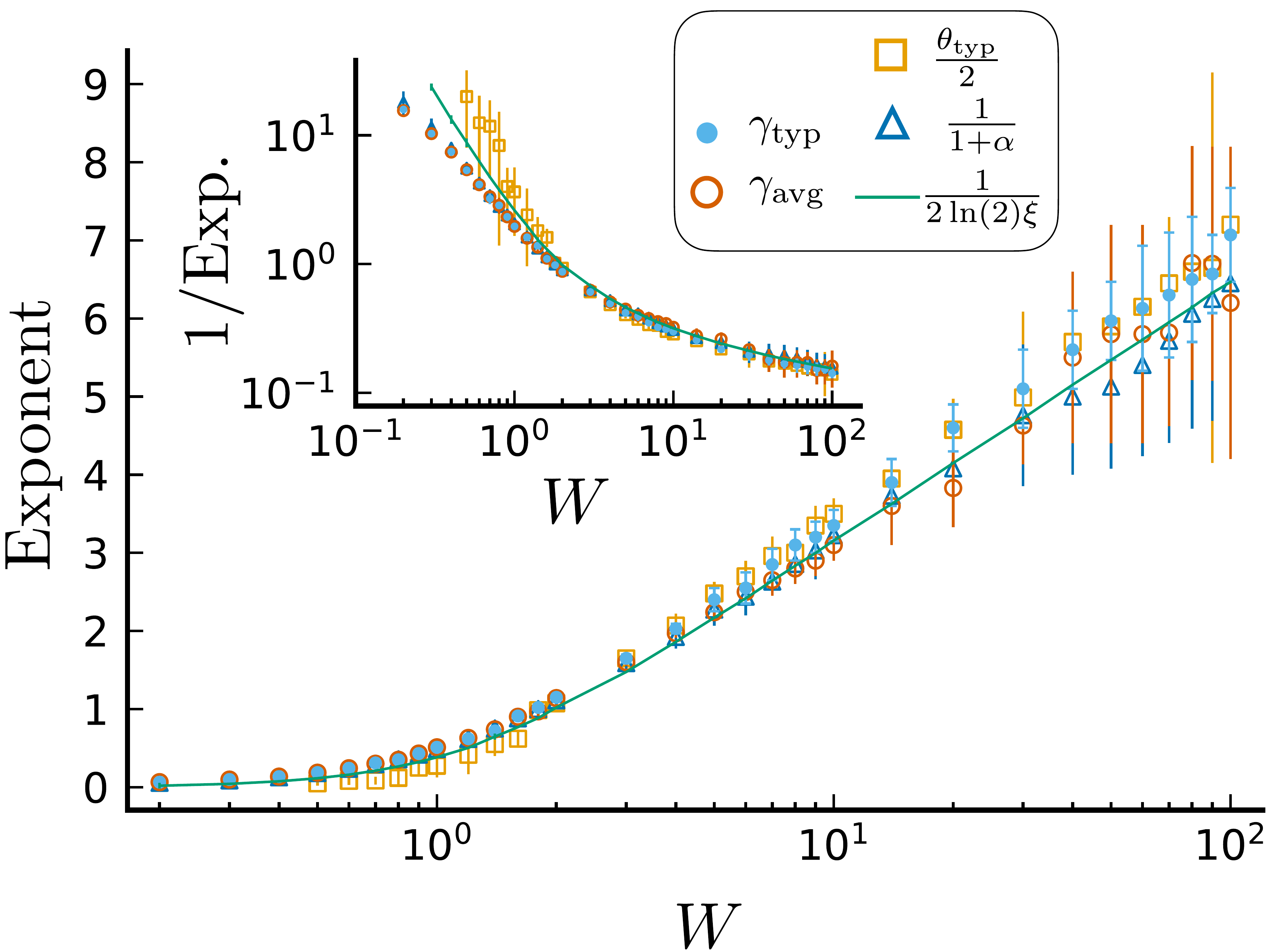}
    \caption{Decay exponents of the minimal deviation ($\gamma_{\avg,\typ}$, Eqs.~\eqref{eq:gammatyp},~\eqref{eq:gammaavg}) and of the spin correlations across the most polarized site ($\theta_{\typ}$, Eq.~\eqref{eq:DefMu}), as a function of the disorder strength $W$. We also show the disorder dependence of the distribution exponent $\alpha$ [Eq.~\eqref{eq:PowerLaw}, Fig.~\ref{fig:alpha}], related to the freezing exponent by $\gamma = 1/(1+\alpha)$ (Eq.~\eqref{eq:gamma_alpha}), and  the toy model result $1/(2\xi\ln2)$ [Eq.~\eqref{eq:gamma_xi}, Fig.~\ref{fig:toy1_f}].  At large disorder, we observe the expected  logarithmic growth with $W$, and the predicted factor of two between weak-link and freezing exponents [Eq.~\eqref{eq:gamma_theta}]. Note that $\gamma_{\rm avg}$ become intractable at strong disorder, yielding very large error bars; for the other error bars see the discussion in Sec.~\ref{sec:SpinFreezing}.}
    \label{fig:Exponents}
\end{figure}

Fig.~\ref{fig:Exponents} shows the disorder dependence of the spin freezing exponents $\gamma_{\typ, \avg}(W)$ controlling the decay of the typical (resp. average) minimal deviation $\delta_{\min}^{\typ,\avg}$, and of the weak-link exponent $\theta_{\typ}(W)$ [Eq.~\eqref{eq:DefMu}] for the decay of the (typical) spin correlations in the direct vicinity of the most polarized site.  
The evaluation of the exponents is limited at weak disorder by the need to reach very large system sizes, and at strong disorder by the need to compute very accurately extremely small deviations $\delta$ (and even smaller correlations $C^{zz}_{\cal LR}$), as mentioned in the previous section. Yet, the predicted Eq.~\eqref{eq:gamma_theta} is indeed verified at strong disorder, and quite interestingly it seems to remain valid down to $W\sim 2$.

\par Further, we plot again the disorder-dependent $\alpha$ exponent, which controls the power-law tails of the deviations distribution, but in a form related to the exponent $\gamma$, as in Eq.~\eqref{eq:gamma_alpha}. The agreement with $\gamma_\typ$ and $\gamma_\avg$ is striking at small disorder (inset of Fig.~\ref{fig:Exponents}), corresponding to the fact that the exponent of the power-law tail indeed controls the decay of the minimal deviation (Sec.~\ref{sec:Frechet_main}). The agreement remains quite good, despite the non-monotonous structures appearing in the distributions which start to affect the results ($W \gtrsim 20$), yielding larger errors and a difference between $\gamma_\typ$, $\gamma_\avg$ and $1/(1+\alpha)$. 

Finally, we note that the disorder-dependence of the exponents is logarithmic at large disorder. This is reminiscent of the disorder-dependence of the inverse localization length, and corresponds to the relation $\gamma \cong (2 \xi \ln 2)^{-1}$ obtained in the simple toy model Eq.~\eqref{eq:gamma_xi}. The agreement between the inverse localization length and the freezing exponent is excellent in Fig.~\ref{fig:Exponents}, down to disorder strengths $W \sim 1-2$, which corresponds to the regime where $\xi$ becomes $O(1)$.

\subsection{Limiting distribution for the extreme polarizations} 
\label{sec:Frechet_main}

We now discuss in detail the distribution of extreme polarizations for finite chains of length $L$.

\subsubsection{Extreme statistics distribution}
\label{sec:Frechet}

\begin{figure*}[tp]
    \centering
    \phantomsubfloat{\label{fig:CollapseDeltaMin_a}}
	\phantomsubfloat{\label{fig:CollapseDeltaMin_b}}
	\phantomsubfloat{\label{fig:CollapseDeltaMin_c}}
	\phantomsubfloat{\label{fig:CollapseDeltaMin_d}}
	\phantomsubfloat{\label{fig:CollapseDeltaMin_e}}
	\phantomsubfloat{\label{fig:CollapseDeltaMin_f}}
	\phantomsubfloat{\label{fig:CollapseDeltaMin_g}}
	\phantomsubfloat{\label{fig:CollapseDeltaMin_h}}
    \includegraphics[width=1.625\columnwidth]{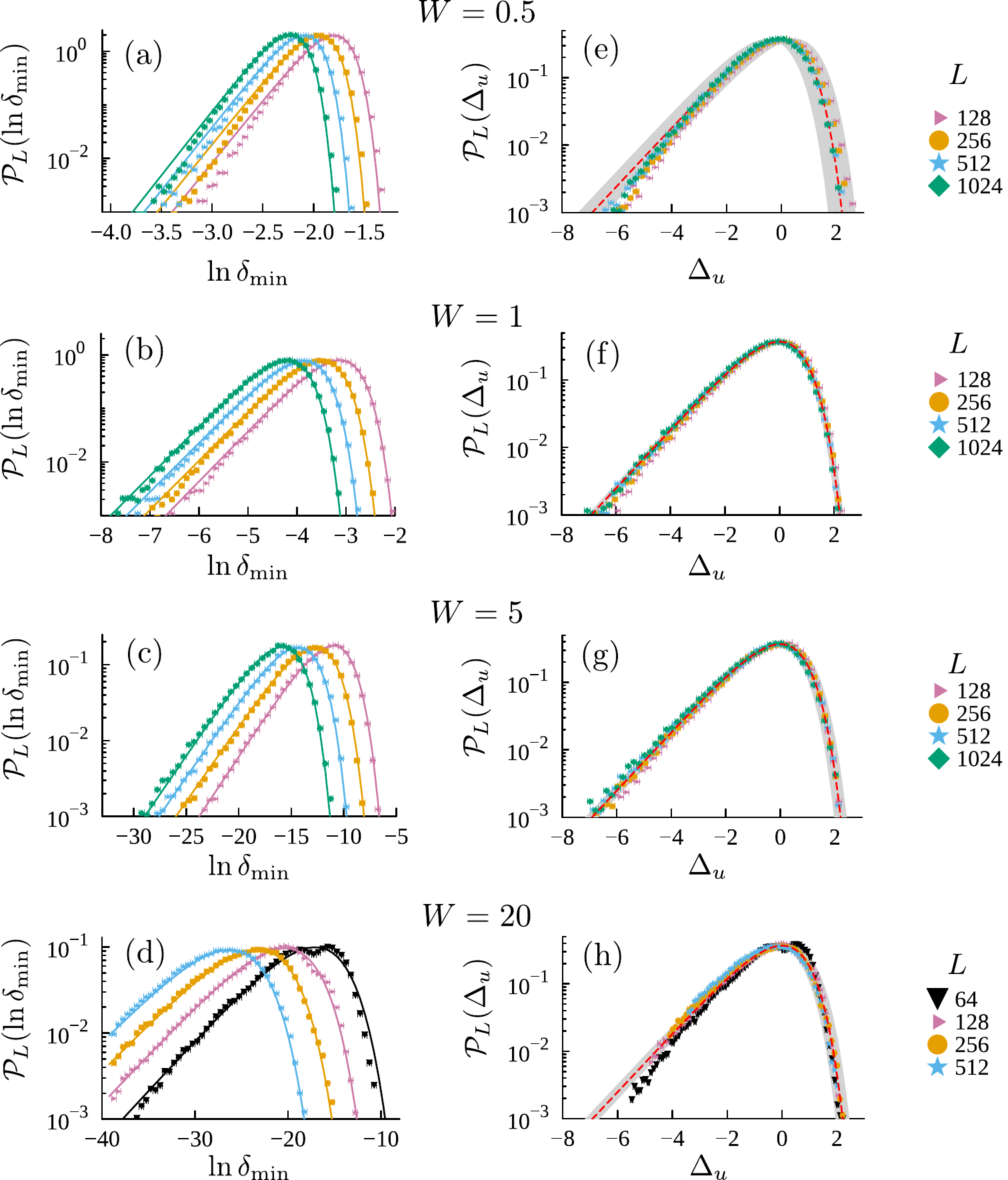}
    \caption{\protect\subref*{fig:CollapseDeltaMin_a} to \protect\subref*{fig:CollapseDeltaMin_d}: ED data for random-field XX chains. Distribution of $\ln\delta_{\min}$ for $10^5$ field realizations for various system sizes. Solid lines are two-parameter fits to the data according to the reflected generalized Gumbel distribution Eq.~\eqref{eq:udist}. \protect\subref*{fig:CollapseDeltaMin_e} to \protect\subref*{fig:CollapseDeltaMin_h}: Collapse of these distributions obtained by defining $\Delta_u= (\alpha+1)(\ln\delta_{\min}- \mu)$, with $\mu$ from Eq.~\eqref{eq:y*}, using the $A$ and $\alpha$ extracted from the fits in Figs.~\protect\subref*{fig:CollapseDeltaMin_a} to \protect\subref*{fig:CollapseDeltaMin_d}. The errors on $\Delta_u$ coming from the uncertainties on $A$ and $\alpha$ are represented by the gray area.
The dashed red line is the same in all four plots and corresponds to the scale-invariant Gumbel law Eq.~\eqref{eq:ydist}. In panels \protect\subref*{fig:CollapseDeltaMin_d} and \protect\subref*{fig:CollapseDeltaMin_h}, we only plot the data for minimal deviations larger than numerical precision. At strong disorder and small sizes, irregular bumps appear (see also Fig.~\ref{fig:DeltaMin_XXvsToy}).}
    \label{fig:CollapseDeltaMin}
\end{figure*}

The Fisher-Tippett-Gnedenko (or extreme value) theorem~\cite{frechet_1927,fisher_limiting_1928,Mises_1936,falk_von_1993,Gnedenko_1943, DeHaan_2006} states that the limiting law controlling the maximum of $L$ independent identically distributed random variables $\{x_i\}_{i= 1, \dots, L}$  can take three forms in the limit of large sizes, depending on the tail of the probability density function $p(x)$ at large $x$. There are three families of limiting distributions which can be gathered in the same \emph{generalized extreme value} (GEV) distribution, controlled by three parameters: the \emph{shape} $s$, the \emph{location} $\mu$ and the \emph{scale} $\sigma$.  The probability density function for the rescaled variable $z = \frac{x_{\max}-\mu}{\sigma}$ is 
\be
    \label{eq:GEV}
    {\cal{P}}(z\,;\,s) = 
    \begin{cases}
        \left(1 + sz\right)^{-(1 + 1/s)}e^{-(1+sz)^{-1/s} } & \text{for } s\neq 0\\
        e^{-(z+e^{-z})} &\text{for } s = 0
    \end{cases}
\ee 
with $(1+sz)>0$. $s = 0$ corresponds to the Gumbel distribution, $s  >0$ Fr\'echet and $s < 0$ the Weibull distribution. 
In the case of a power-law tail of the parent distribution at large $x$,
\begin{equation}
    p(x) = A x^{-\beta-1},\quad \beta > 0,
\end{equation}
a Fr\'echet law~\cite{frechet_1927} is expected for the distribution of the rescaled maximum $z = x_{\max}\left(\frac{AL}{\beta+1}\right)^{-\frac{1}{\beta}}$
\be
\label{eq:frechetdist}
{\cal{P}}_{L}\left(z\right) \rightarrow \beta z^{-\beta-1}\exp\left(-z^{-\beta}\right)
\ee
As numerically  observed in Sec.~\ref{sec:Mdistrib}, we expect that the deviations $\delta$ from perfect polarization are distributed according to a power-law $\mathcal{P}(\delta)\sim \delta^\alpha$ when $\delta\to 0$ (i.e. $m_i = |\langle S_i^{z}\rangle|\rightarrow 1/2$), see Eq.~\eqref{eq:PowerLaw} and Fig.~\ref{fig:LnDeviationsHistogram}, with the disorder-dependent exponent $\alpha(W)>-1$ shown in Fig.~\ref{fig:alpha}. If this is indeed the case, then the minimal deviation for chains of length $L$, computed as follows
\be
\delta_{\min} = \frac{1}{\max_{i=1,\ldots,L} \left(\frac{1}{\delta_i}\right)},
\ee
is expected to be distributed according to the Fr\'echet-related law \footnote{An alternative derivation of this result is given in~\cite{SM} Sec.~S3, using the fact that the distribution for $m_i$ is upper-bounded.}:
\be
\label{eq:dmindist}
{\cal{P}}_{L}(\delta_{\min}) \rightarrow AL \delta_{\min}^{\alpha} \exp\left({-\frac{AL}{\alpha+1}\delta_{\min}^{\alpha+1}}\right).
\ee
Changing variable to the logarithm of the minimal deviation $u = \ln \delta_{\min}$, we get the (reflected) generalized Gumbel law

\be
	{\cal{P}}_L(u) \rightarrow AL \exp\left[(\alpha+1) u -\frac{AL}{\alpha+1}{\rm{e}}^{(\alpha+1)u}\right],
	\label{eq:udist}
\ee
with scale  $\sigma = (\alpha+1)^{-1}$, and location  $\mu$ given by
\be
	\mu = -\frac{1}{\alpha+1} \ln\left(\frac{AL}{\alpha+1}\right).
	\label{eq:y*}
\ee
This non-trivial dependence Eq.~\eqref{eq:udist} is checked for the numerical data using two-parameter fits in Fig.~\ref{fig:CollapseDeltaMin}. The location Eq.~\eqref{eq:y*} directly gives the maximum of the distribution, yielding another estimate for the typical deviation:
\be
\label{eq:dminFrechet}
\delta_{\min}^{\typ}(L)\approx \left(\frac{A}{\alpha+1}\, L\right)^{-\frac{1}{\alpha+1}},
\ee 
which nicely matches Eq.~\eqref{eq:deltamin}.
Furthermore, the change of variables $\Delta_u = (\alpha+1)(\ln \delta_{\rm min}-\mu)$, yields the scale-invariant (reflected) Gumbel distribution
\be
    {\cal{P}}_{L}(\Delta_u)\approx  \exp\left[\Delta_u-\exp(\Delta_u)\right],
    \label{eq:ydist}
\ee
perfectly illustrated on the right side of Fig.~\ref{fig:CollapseDeltaMin}. It is important to note that this Gumbel distribution for $\ln\delta_{\rm min}$  corresponds to a Fr\'echet law for the inverse minimal deviations.

On the left side, fits to the  law Eq.~\eqref{eq:udist} are performed using $A$ and $\alpha$ as free parameters. The collapse scale-invariant form Eq.~\eqref{eq:ydist} is then obtained using $A$ and $\alpha$ averaged over the various system sizes (errors are shown by the gray area, see also Sec.~S3 in~\cite{SM}). At first sight the agreement with the Gumbel distributions is very good. Correspondingly, the extracted values of $A$ and $\alpha$ can be directly related to the amplitude and the exponent of the decay of $\delta_{\min}^{\typ}$ in Eq.~\eqref{eq:dminFrechet}. We have checked (\cite{SM} Sec.~S3) that the resulting values of $\alpha$ agree within errors with the power-law tail of the deviations (Fig.~\ref{fig:LnDeviationsHistogram}) and with the freezing exponent $\gamma$. Recall that only the tails of the distributions ${\cal{P}}(\delta)$ follow a pure power-law, thus the amplitudes $A$ extracted from the fits in Figs.~\ref{fig:CollapseDeltaMin_a} to~\ref{fig:CollapseDeltaMin_d} slightly differ from theoretical expectations coming from a perfect power-law: $A \approx  (\alpha +1 ) 2^{(\alpha+1)}$.

\subsubsection{Small corrections to the limiting distributions}
\label{sec:corr}
There are several sources of small corrections. First, we should recall that the extreme value distributions are only the limiting distributions, and finite-size corrections should be expected (see e.g.~\cite{Kovacs_2021}). Second, it is important to say that the deviations $\delta_i$ are not fully independent random variables, and therefore one might expect some small differences with the limiting extreme value theory distributions. In fact, the chain breaking mechanism tells us that the extreme polarizations occur in clusters and are thus correlated within each clusters, see Fig.~\ref{fig:toy1_b}. Fortunately, these correlations are found to be weak (Sec.~S3 in~\cite{SM}): the spatial correlations of the deviations $\delta_i$, averaged over disorder, decay exponentially with a quite small correlation length $\zeta$, remaining below order of ten sites even at weak disorder. In such a case, it is known~\cite{Majumdar_2020} that one can predict the extreme statistics distribution based on the distribution function of the local minima for a system divided in blocks of length $\zeta$. Here, we expect the local minima to still follow a power-law distribution, and the extreme value distribution must describe the statistics of the global minima, as clearly visible in Fig.~\ref{fig:CollapseDeltaMin}.

\par Nevertheless, we observe some small discrepancies. First, at weak disorder, the left tail is not well fitted by the limiting law; as a consequence the collapse in Fig.~\ref{fig:CollapseDeltaMin_d} has relatively large errors. We suspect this small skew to come from the slow convergence of the power-law tail with the system size in this regime. Indeed, the extreme value description seems to improve when both $L$ and disorder strength grow.
However,  additional (non-monotonous) structures appear for very strong disorder: note that they are already visible in Fig.~\ref{fig:LnDeviationsHistogram_b} for ${\cal{P}}(\ln \delta_i)$. On the other hand, it is important to say that as the disorder increases, the tail of the power law remains the correct description for very small deviations. For example, data at $W = 20$ in Fig.~\ref{fig:CollapseDeltaMin_d} show that the larger system sizes are still well described by the power-law tail at very small deviations, but this will gradually degrade when $W$ becomes very large, as more and more bumpy structures would develop. Interestingly, these features have a simple microscopic origin very well captured by the toy model, which will be discussed below, in Sec.~\ref{sec:ToyModel}.

\subsubsection{Connections with the random transverse-field Ising chain }
Overall, the histograms of extreme polarizations ${\cal{P}}(\delta_{\rm min})$ are fairly well described by the extreme value theory over the entire localized regime (Fig.~\ref{fig:CollapseDeltaMin}), following the distribution Eq.~\eqref{eq:dmindist} coming from a Fréchet distribution which is essentially controlled by the disorder-dependent exponent $\alpha$, primarily governing the tails of the parent distributions ${\cal{P}}(\delta)\sim \delta^{\alpha}$. While describing a totally different physical effect, a similar Fr\'echet law also characterizes the distribution of the lowest finite-size energy gap $\epsilon_{\rm min}(L)$ of the random transverse-field Ising chain model (TFIM)~\cite{fisher1998,Juhasz_2006,Kovacs_2021,Kao_2022}. In that case, the relevant exponent is the dynamical exponent $z$ which controls the singular distribution of renormalized energy scales in the disordered regime~\cite{fisher_critical_1995} such that ${\cal{P}}(\epsilon)\sim \epsilon^{-1+1/z}$, thus yielding  $\epsilon_{\rm min}(L)\sim L^{-z}$~\footnote{Agreements and small deviations to Fr\'echet have been thoroughly discussed in Ref.~\onlinecite{Kovacs_2021}.}. Interestingly, one can make a formal analogy between the finite-size gap vanishing in the random TFIM and the chain-breaking occurring in the many-body Anderson insulator. The freezing exponent $\gamma$ is the analog of $z$, but  they remarkably behave in opposite ways with their localization lengths: $\gamma\sim 1/\xi$ at strong disorder, in stark contrast with the divergence $z\sim \xi$ in the vicinity of the critical point of the random TFIM.

\section{Non-interacting toy model}
\label{sec:ToyModel}

\begin{figure*}
    \centering	
    \includegraphics[width=\textwidth]{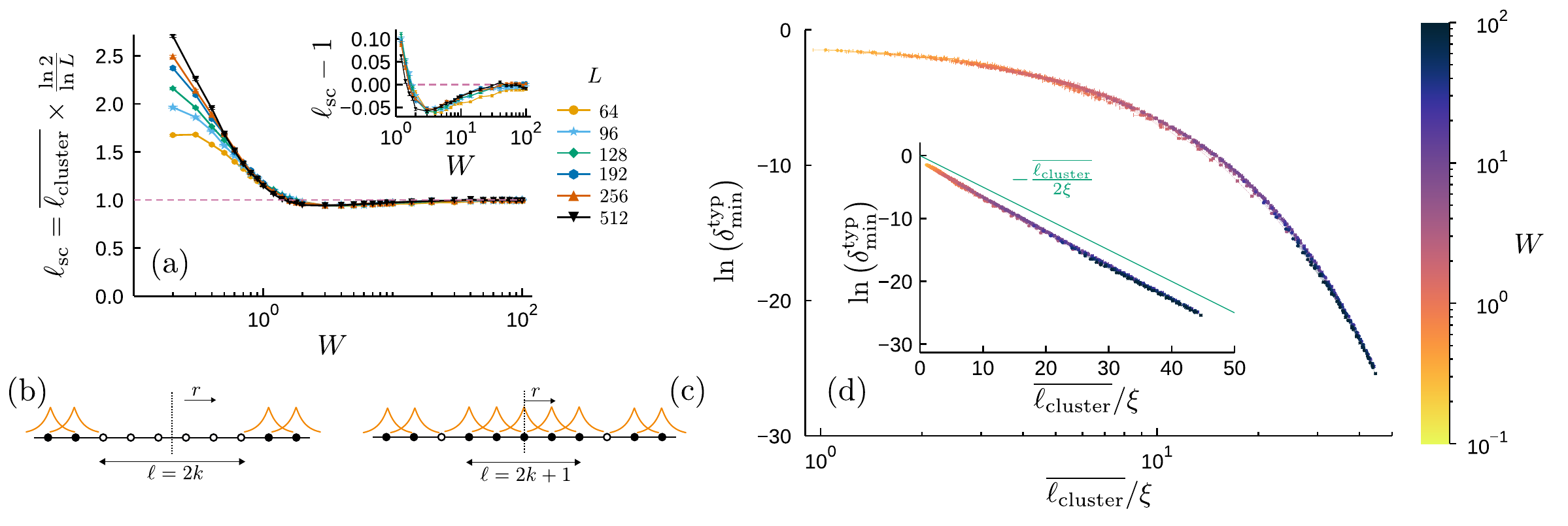}
    \phantomsubfloat{\label{fig:PlCluster}}
    \phantomsubfloat{\label{fig:ToySketchb}}
    \phantomsubfloat{\label{fig:ToySketchc}}
    \phantomsubfloat{\label{fig:SDDeltaMinC}}
    \vskip -1cm
    \caption{Validity of the toy model for the description of the deviation $\delta_{\min}$ of the most polarized site $i^{\star}$. \protect\subref*{fig:PlCluster} Average length $\overline{\ell_{\cluster}}$ of the cluster containing the most polarized site, as determined by the occupation of this site and its neighbors (see text), and rescaled with the expected size dependence at strong disorder, $\ln L /\ln 2$. The inset is a zoom on the difference of the rescaled length to one.  \protect\subref*{fig:ToySketchb} Example of an even, empty, isolated cluster in the toy model. \protect\subref*{fig:ToySketchc} Example of an odd, occupied, non-isolated cluster in the toy model. \protect\subref*{fig:SDDeltaMinC} ED data for the power-law decay of the typical value of the minimal deviation, rescaled with the localization length $\xi(W)$ and the average cluster length, for sizes $L \geq 12$. Inset: same data but in linear scale. The green line is a guide to the eye corresponding to Eq.~\eqref{eq:toymin} with a zero shift. Compare to Fig.~\ref{fig:SizeDepDeltaMin}. }
    \label{fig:CollapseDeltaMinTyp}
\end{figure*}

As introduced in Sec.~\ref{sec:CB}, it was suggested in Refs.~\onlinecite{Dupont_2019, Laflorencie_2020} that the algebraic vanishing of the minimal deviation with the system size 
$\delta_{\rm{min}}(L) \sim L^{-\gamma}$ can be readily explained by a simple toy model, in principle valid at strong disorder. Supposing that each orbital $\phi_m$ is localized around a well-defined localization center $i_0^m$, see Fig.~\ref{fig:toy1_d}, it is further assumed  that its envelope decays exponentially and symmetrically around that center, with a site-independent localization length $\xi$, such that $\left|\phi_m(i)\right|^2\sim \exp(-|i-i_0^m|/\xi)$.

As we have sketched in Sec.~\ref{sec:CB}, this toy model is able to explain the main feature of the spin freezing by relating the typical deviation $\delta_{\min}^{\typ}$ at half-filling to the average of the maximal sequence $\ell_{\max}$ of neighboring occupied or empty orbitals in real space, Eq.~\eqref{eq:deltamin_lmax}, resulting in a chain-breaking mechanism  controlled by the freezing exponent $\gamma \sim (2\xi \ln2)^{-1}$. In this section we want to clarify and nuance the toy model results and draw further consequences.

\subsection{Scaling for the spin freezing}

\subsubsection{Localization length}
The unique parameter of the toy model is the $W$-dependent localization length $\xi(W)$, where the energy dependence has been integrated over the single-particle density of states 
(see Sec.~S2 in~\cite{SM} for the computation  of $\xi(E,W)$ from the Lyapunov exponent~\cite{Kramer_1993, Crisanti_1993,Comtet_2013}).
This energy-averaged localization length is known to diverge at weak disorder $ \sim 1/W^2$~\cite{Kappus_1981,Kramer_1993}, while for strong randomness a simple perturbative expansion of the wavefunction around its localization center shows that $\xi^{-1} \sim 2 \ln(W)$ \cite{Laflorencie_2022}. Therefore, the simple Ansatz formula~\cite{Potter_2015} which combines both weak and strong disorder limits
\begin{equation}
	\label{eq:XiFit}
	\xi =\frac{1}{ \ln\left[1 + \left({W}/{W_0}\right)^2\right]}
\end{equation}
nicely fits the bill, as shown in Fig.~\ref{fig:toy1_g} where we see that Eq.~\eqref{eq:XiFit} with $W_0 \sim 1.2$ captures extremely well the exact numerics for $\xi(W)$.

As introduced in Sec.~\ref{sec:CB}, the a priori simplistic toy model provides a remarkably realistic description of the many-body Anderson insulator. In particular, the simple expression for the freezing exponent $ \gamma \sim (2\xi\ln2)^{-1}$ remains valid over a very broad range of randomness, and only starts to deviate typically below $W \sim 2$, as clearly shown in Fig.~\ref{fig:Exponents}. Nevertheless, in what follows, we are going to see that the extreme polarization scaling derived within the simple toy-model framework, $\delta_{\min}(L)\sim \exp\left(-\frac{\ell_{\max}}{2\xi}\right)$, can be extended to weaker disorder strengths, provided the fact that the maximal sequence $\ell_{\max}$ is replaced by ${\overline{\ell_{\rm cluster}}}$, the average length of the cluster hosting the most polarized spin.

\subsubsection{Cluster length}
Let us first define the cluster length $\ell_{\cluster}$ for any given sample as the size of the region surrounding the most polarized site in which the magnetization does not change sign (see green region in Fig.~\ref{fig:toy1_a}). The motivation for this definition is twofold: (i) it holds even at weak disorder, in particular when the localization length is large and a one-to-one correspondence between sites and orbitals becomes ill-defined, and (ii) it remains valid in the presence of finite interactions, namely for the MBL problem discussed in Sec.~\ref{sec:MBL}.

We have numerically computed the disorder average cluster size, and its rescaled form $\ell_{\rm sc}=\lc/{\ell_{\max}}$ is shown in Fig.~\ref{fig:PlCluster}. At strong disorder, we expect the cluster lengths to be controlled by $\ell_{\max}= \ln L / \ln 2$. This is indeed what is observed in Fig.~\ref{fig:PlCluster} for a surprisingly wide range of disorder strengths, down to $W^* \sim 1.5$, thus giving a rough estimate for the range of validity of the toy model. Remarkably, for $W \leq W^*$ the average cluster length becomes significantly larger than ${\ell_{\max}}$ and strong finite-size corrections start to appear, while at intermediate values of $W$ the average cluster length is slightly below $\ell_{\max}$. This non-monotonous behavior, best visible in the inset of Fig.~\ref{fig:PlCluster}, results from the competition between two effects. At intermediate disorder  (typically $1<W<10$) spatial fluctuations in the localization lengths can  lead to local configurations for which the most polarized site may belong to a cluster slightly smaller than $\ell_{\max}$. On the other hand at smaller $W$, where the notion of localization center start to become fuzzier, sites that should be normally associated to an empty orbital can have slightly more than half-occupation, thus creating wider and wider clusters, as clearly shown in Fig.~\ref{fig:PlCluster}. Correspondingly, the maximal number of sites with the same sign of the magnetization remains everywhere larger or equal to $\overline{\ell_{\cluster}}$ and can become extremely large at weak disorder, deviating very strongly from the toy model value ${\ell_{\max}}$.

\begin{figure*}
    \centering
    \phantomsubfloat{\label{fig:LnDelta_a}}
    \phantomsubfloat{\label{fig:LnDelta_b}}
    \phantomsubfloat{\label{fig:LnDeltaMin_c}}
    \phantomsubfloat{\label{fig:LnDeltaMin_d}}
    \phantomsubfloat{\label{fig:LnDeltaMin_e}}
    \phantomsubfloat{\label{fig:LnDeltaMin_f}}
    \includegraphics[width=\textwidth]{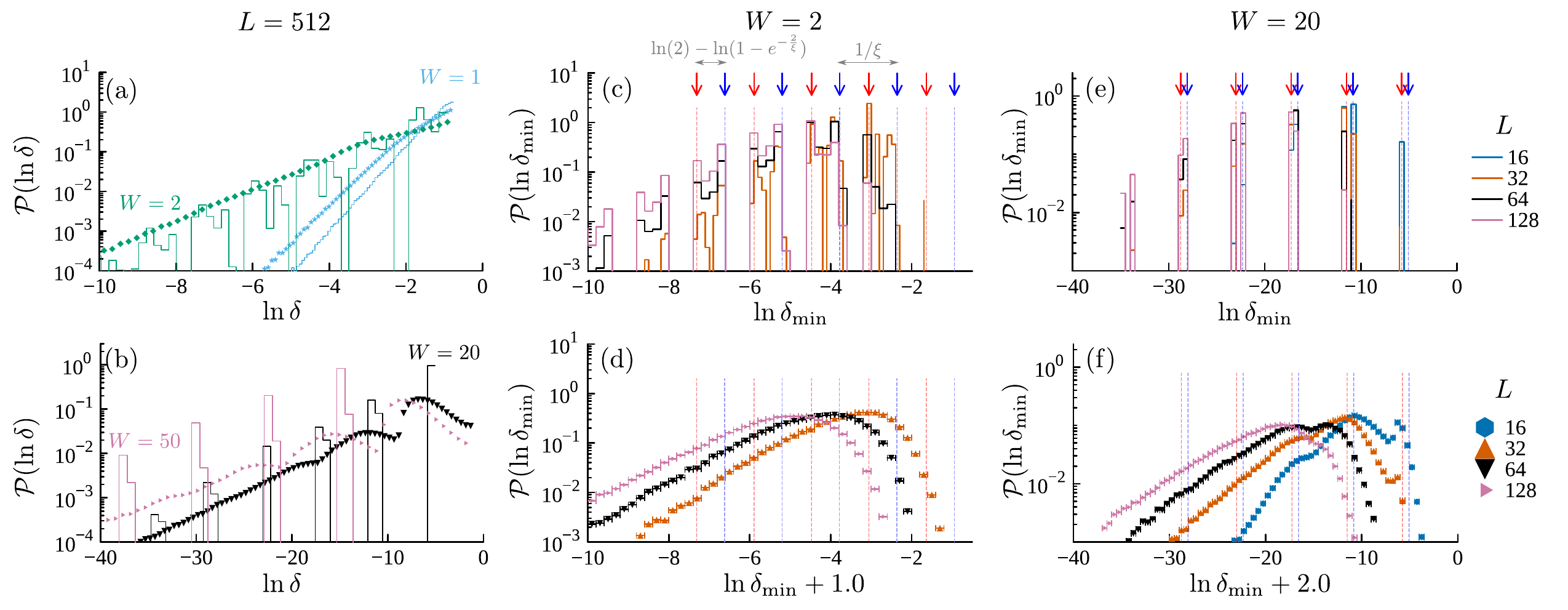}
    \caption{\protect\subref*{fig:LnDelta_a} and~\protect\subref*{fig:LnDelta_b} Distribution of the deviations in the XX chain (symbols) at various disorder strengths $W$ and in the toy model (stepped histograms) at the corresponding localization length $\xi(W)$, for chains of 512 sites. \protect\subref*{fig:LnDeltaMin_c} Distribution of the minimal deviations in the toy model at $W = 2$. The arrows indicate the location of the peaks corresponding to (in blue) odd length clusters in Eq.~\eqref{eq:upperodd} and (in red) even-length clusters in Eq.~\eqref{eq:lowereven}. In gray, we indicate the expression for the spacing between the peaks. \protect\subref*{fig:LnDeltaMin_d} 
 Corresponding ED distributions in the XX chain, shifted by $\ln D_W =  1$. \protect\subref*{fig:LnDeltaMin_e}  Distribution of the minimal distributions in the toy model at $W = 20$.  \protect\subref*{fig:LnDeltaMin_f} Corresponding ED distributions in the XX chain, shifted by $\ln D_W = 2$.}
    \label{fig:DeltaMin_XXvsToy}
\end{figure*}

%
\subsubsection{Scaling plot and data collapse}

With this in hands, we can now re-write the toy model result Eq.~\eqref{eq:deltamin_lmax} as an expression that can be tested at large scales for the XX chain: 
\be 
	\ln(\delta_{\min}^{\typ})  = -\frac{\lc}{2\xi} - \ln D_W,
	\label{eq:toymin}
\ee
where $D_W$ is a disorder-dependent constant. Using the data from Fig.~\ref{fig:SizeDepDeltaMin} and plotting it as a function of the ratio of the mean cluster length to the localization length, in Fig.~\ref{fig:SDDeltaMinC}, we observe and excellent collapse of the numerical data.  The fact that a good collapse is obtained even when neglecting $\ln D_W$ suggests that this correction has a weak disorder dependence. 

\par In fact, at weak to intermediate disorders ($W \lesssim 10$), where $\lc$ differs from the simple toy model prediction, the collapse is much better than when using $\ln(L)/\xi$ as a scaling parameter (Sec.~S4 in~\cite{SM}). Only small deviations to a perfect collapse can be observed for large sizes for these disorder strengths. This means that the deviation of $\lc$ from the expected $\ln(L)/\ln(2)$, observed in Fig.~\ref{fig:PlCluster}, fully accounts for the deviation of $\gamma$ from $1/(2\xi\ln2)$ observed in Fig.~\ref{fig:Exponents}. Thus, Fig.~\ref{fig:Exponents} and further Fig.~\ref{fig:CollapseDeltaMinTyp} show that the toy model yields an excellent description of the chain breaking mechanism for the typical minimal deviation.

We now describe how the toy model helps to explain features of the minimal deviation distributions not captured by extreme value theory, in particular at strong disorder and small sizes.

\subsection{Distributions}
\label{sec:TM_Dist}

The toy model allows to exactly obtain some simple results. Indeed, one can easily compute the deviation at a distance $r$ from the center of an isolated cluster of empty orbitals surrounded by occupied orbitals, as in Fig.~\ref{fig:ToySketchb}, or of a cluster of occupied orbitals only separated from other clusters by two empty orbitals, as in Fig.~\ref{fig:ToySketchc}.  These simple calculations directly yield several results for the distribution of the deviations and of the minimal deviations in the toy model. 

\subsubsection{Parent distribution ${\cal{P}}(\delta)$}
First, the power-law tail is easily recovered (Sec.~S4 in~\cite{SM}): remembering that at half-filling the probability of having a cluster of length $\ell$ is $\mathcal{P}(\ell) \propto 2^{-\ell}$ we can roughly estimate the probability of a deviation $\delta$ and find that
\begin{equation}
	\mathcal{P}_L(\ln(\delta)) \propto 2^{-2k} \sim 2^{2\xi\ln(\delta)} = e^{\frac{\ln(\delta)}{\gamma}}
\end{equation}
in agreement with Eq.~\eqref{eq:PowerLaw}.

\par Second, we find that there must be peaks in the distribution around locations corresponding to integer values of the cluster length $\ell$: $\ln\delta \sim -\ell/(2\xi)$. This is what we show in Figs.~\ref{fig:LnDelta_a} and~\subref*{fig:LnDelta_b}. As can be seen in Fig.~\ref{fig:LnDelta_a}, at weak disorder, the peaks merge into a single distribution in the toy model. The exponent is clearly different from the XX chain exponent, corresponding to the difference between the toy model prediction $\gamma = (2\xi\ln2)^{-1}$ and $\gamma_{\typ}$ shown in Fig.~\ref{fig:Exponents}. This comes from the difference between $\overline{\ell_{\cluster}}$ and $\ell_{\max}$ discussed in the previous part. In contrast, at strong disorder, the peaks are well separated in the toy model, as visible in Fig.~\ref{fig:LnDelta_b}. Recall that in the toy model picture, the deviations are solely controlled by the distribution of cluster lengths, while the effect of the distribution of the localization lengths is neglected. In the XX chain, the fluctuations of the localization lengths smooth out the peaks, resulting in the bumpy features noted in Fig.~\ref{fig:LnDeviationsHistogram_b}.

\subsubsection{Extreme value distribution ${\cal{P}}(\delta_{\rm min})$}
Now, let us discuss the distribution of the minimal deviations. The rough estimate Eq.~\eqref{eq:deltamin_lmax} of the relation between the minimal deviation and the maximal cluster length can be refined.  Considering the situation depicted in Fig.~\ref{fig:ToySketchb}, we get an upper bound for the minimal deviation in the toy model (see Sec.~S4 in~\cite{SM} for detailed calculations). For an even-length cluster, we have
\begin{equation}
    \label{eq:uppereven}
    \delta_{\min}(\ell) = e^{-\frac{\ell}{2\xi}}, \quad  \ell = 2k, 
\end{equation}
while for an odd-length cluster, we obtain
\begin{equation}
    \label{eq:upperodd}
    \delta_{\min}(\ell) = \frac{e^{-\frac{\ell}{2\xi}}}{\cosh\left(\frac{1}{2\xi}\right)},  \quad  \ell = 2k-1. 
\end{equation}

Thus, well-defined peaks must also appear in the distribution of the logarithm of the minimal deviations in the toy model, as shown in Figs.~\ref{fig:LnDeltaMin_c} and~\subref*{fig:LnDeltaMin_e}. The distance between peaks coming from clusters of the same parity is controlled by the inverse localization length, in agreement with Eqs.~\eqref{eq:uppereven} and~\eqref{eq:upperodd}. We note that the location of the peak due to an even-sized cluster of size $\ell = 2k$ in Eq.~\eqref{eq:uppereven} is always slightly smaller than that coming from a cluster of size $\ell = 2k-1$. Only the latter is depicted (in blue arrows) in Fig.~\ref{fig:DeltaMin_XXvsToy}. 
\par In contrast, the situation in Fig.~\ref{fig:ToySketchc} yields an estimate for a lower bound for the minimal deviation in a cluster of length $\ell$. For an even-sized cluster, we have
\begin{equation}
    \label{eq:lowereven}
    \delta_{\min}^{\mathrm{lower}} (\ell) = (1-e^{-\frac{1}{\xi}})e^{-\frac{\ell}{2\xi}},
\end{equation}
a result depicted by red arrows in Figs.~\ref{fig:LnDeltaMin_c} and~\subref*{fig:LnDeltaMin_e}. We show in Figs.~\ref{fig:LnDeltaMin_c} and~\subref*{fig:LnDeltaMin_e} that the toy model distributions are indeed well-captured by this simple description. 

\par Finally, in the minimal deviation distribution, the relative height of the peaks is directly related to the distribution of the maximal cluster lengths $\mathcal{P}^{\max}_L\left(\ell\right)$ via
\begin{equation}
	\mathcal{P}_L \left(\ln\left(\delta_0^{(\ell)}\right)\right) = \frac{\xi}{2} \mathcal{P}^{\max}_L\left(\ell\right).
\end{equation}
Thus, as the system size increases, the probability weight shifts from small negative peaks towards larger negative values, but the location of the peaks does not change, as clearly depicted in Figs.~\ref{fig:LnDeltaMin_c} and~\subref*{fig:LnDeltaMin_e}. 
\par In the XX chain, we recover all these effects for large deviations, with further corrections due to the distribution in localization lengths (Figs.~\ref{fig:LnDeltaMin_d} and~\ref{fig:LnDeltaMin_f}). Since the prefactor $D_W$ of the exponential decay of the minimal deviation is controlled also by the localization lengths distribution, there is a shift between the locations of the peaks in the toy model and the XX chain, which we approximate by shifting the data in Figs.~\ref{fig:LnDeltaMin_d} and~\ref{fig:LnDeltaMin_f} by the shift $D_W$ obtained from the typical minimal deviation in Fig.~\ref{fig:SDDeltaMinC}. Thus, the simple toy model provides a good rule-of-thumb for the location and height of the bumpy features not captured by the extreme value distributions for small to intermediate system sizes and large $W$.

\section{Consequences for the MBL problem}
\label{sec:MBL}
We now turn to finite interactions, in the middle of the many-body spectrum, a case which has generated a huge amount of theoretical and experimental activity during the last decade: for recent reviews see~\cite{alet_many-body_2018,abanin_many-body_2019}. While the status of MBL was thought to be well understood in 1D, a new debate has recently emerged~\cite{Panda_2019,morningstar2021avalanches,PhysRevB.105.224203}. 
Indeed, the broken ergodicity, numerically observed for the random-field Heisenberg chain model
\be
{\cal{H}}_{\rm Heisenberg}=\sum_i\left({\vec{S}}_{i}\cdot {\vec{S}}_{i+1}-h_iS_i^z\right),
\label{eq:H}
\ee
at a critical disorder strength $W_c$  in a relatively large range $3.5\le W_c\le 5.5$~\cite{luitz_many-body_2015,doggen_many-body_2018,PhysRevB.97.201105,Laflorencie_2020,chanda_time_2020}, has been argued to be only a mere finite-size phenomenon, with the "true" MBL transition occurring at much larger disorder strengths $W_c>10$ or even $W_c>20$~\cite{PhysRevB.106.L020202}. Incidentally, this would correspond to a tiny non-interacting average localization length $\xi\approx 0.2$.
In light of these recent developments, the objective of this section is to open the toolbox of extreme value statistics in order to reexamine the ergodicity breaking transition for the celebrated random-field Heisenberg chain Hamiltonian Eq.~\eqref{eq:H}. This part is organized as follows. In Sec.~\ref{sec:ITM} we first start from the non-interacting Anderson limit, and rewrite the interacting model in the Anderson basis from which one can derive a meaningful interacting toy model which helps us to address the fate of the freezing mechanism in the presence of interactions. We then use state-of-the-art numerical simulations (shift-invert ED up to $L=22$ sites) in 
Sec.~\ref{sec:NUMMBL} to quantitatively explore the freezing and compare Heisenberg with XX data. Sec.~\ref{sec:EVSMBL} is then devoted to the EVS analysis. Building on the Anderson insulator results of the previous sections, a direct comparison of the distributions of extreme polarizations for the interacting case turns out to be very instructive, thanks to the quantitative tool provided by the Kullbach-Leibler divergences~\cite{kullback1951information}.

\subsection{Interacting toy model}
\label{sec:ITM}
\subsubsection{XXZ model in the Anderson basis}
Adding interactions to the XX chain Hamiltonian Eq.~\eqref{eq:XX} is straightforward, yielding the well-known XXZ model 
\be
    \mathcal{H}_{\rm xxz}={\cal{H}}_{\rm xx}+ V_{\rm z},
    \label{eq:XXZ}
\ee
where ${\cal H}_{\rm xx}$ describes free fermions, see Eq.~\eqref{eq:XX}, and the interacting part is given (up to an irrelevant constant)  by
\be
V_{\rm z}=\Delta\sum_{i=1}^{L}S_i^z S^z_{i+1}=\Delta
\sum_{i=1}^{L}n_i n_{i+1}.
\label{eq:Vz}
\ee
Building on the previous result Eq.~\eqref{eq:diag}, ${\cal H}_{\rm xx}$
is diagonalized by the canonical transformation $b_m=\sum_{i=1}^{L}\phi_m(i)c_i$.
In this Anderson basis of localized orbitals $\{\phi_m\}$, the interaction term Eq.~\eqref{eq:Vz} reads
\be
V_{\rm z}=\sum_{l,m,p,q}V_{lmpq}b^{\dagger}_{l}b^{\vphantom\dagger}_{m}b_{p}^{{\dagger}}b_{q}^{\vphantom{\dagger}},
\label{eq:VzA}
\ee
with matrix elements~\cite{Prelovsek_2018,Thomson_2018,Tomasi_2019}
\be
V_{lmpq}=\Delta\sum_{i=1}^{L}\phi_l(i)\phi_m(i)\phi_p(i+1)\phi_q(i+1).
\label{eq:Vanderson}
\ee
One can decompose the interaction Hamiltonian Eq.~\eqref{eq:VzA} in 4 parts, introducing the density operator in the Anderson basis $n_l=b^{\dagger}_{l}b^{\vphantom\dagger}_{l}$
\bea
V_{\rm z}&=&\sum_m{\cal V}^{(1)}_{m}n_{m}+\sum_{l\neq m}{\cal V}^{(2)}_{l,m}n_l n_m\\
&+&\sum_{l\neq m\neq p}{\cal V}^{(3)}_{l,m,p}n_l b^{\dagger}_{m}b^{\vphantom\dagger}_{p}
+\sum_{l\neq m\neq p\neq q}{\cal V}^{(4)}_{l,m,p,q}b^{\dagger}_{l}b^{\vphantom\dagger}_{m}b^{\dagger}_{p}b^{\vphantom\dagger}_{q}.\nonumber
\label{eq:all_terms}
\eea
The first two terms ${\cal V}^{(1,2)}$
 are diagonal, and can be interpreted as a first approximation for the celebrated $l$-bit Hamiltonian~\cite{Serbyn_2013,Huse_2014,Ros_2015,Monthus_2016,Rademaker_2017,pancotti_almost_2018}, while ${\cal V}^{(3,4)}$ are off-diagonal.

\subsubsection{Analytical expression for the toy model}
The interacting XXZ Hamiltonian Eq.~\eqref{eq:XXZ} in a random magnetic field takes the following toy model form, expected to be valid for strong disorder. Further assuming that all Anderson orbitals are localized with a single well-defined localization center, and with a unique localization length $\xi$ given by Eq.~\eqref{eq:XiFit}, we get (see supplemental material of Ref.~\onlinecite{Laflorencie_2020})
\bea
{\cal{H}}_{\rm xxz}&=&\sum_{m=1}^{L}\Bigl[\left({\cal{E}}_m+{\cal{V}}^{(1)}\right)n_m+\sum_{r\ge 1}J^{(2)}_{r}n_m n_{m+r}\nonumber\\
&+&\sum_{1\le r< r'}J^{(3)}_{r,r'}n_m b^{\dagger}_{m+r'}b^{\vphantom\dagger}_{m+r}\nonumber\\
&+&\sum_{1\le r< r'< r''}J^{(4)}_{r,r',r''}b^{\dagger}_{m}b^{\vphantom\dagger}_{m+r}b^{\dagger}_{m+r''}b^{\vphantom\dagger}_{m+r'}\Bigr],\label{eq:simple}
\eea
where ${\cal{E}}_m$ are single-particle~energies, ${\cal{V}}^{(1)}\approx\Delta/W^2$, and the diagonal "density-density" term $J^{(2)}_{r}\sim \Delta\exp\left[-{(r-1)}/{\xi}\right]$. The off-diagonal contributions $J^{(3,4)}$ also vanish exponentially with the inter-orbital distances~\cite{Laflorencie_2020}, the decay being again controlled by the non-interacting localization length $\xi$.
We have deliberatly ignored the non-trivial random sign structure of the above terms, expected from Eq.~\eqref{eq:Vanderson}, which is undoubtedly frustrating. However, since we are working at high energy, the sign of the couplings should not  be relevant in the infinite temperature limit~\cite{Lin_2018}.

\subsubsection{Perturbative arguments for the stability of  the freezing process}
The pressing question here is: how the off-diagonal terms in the above Hamiltonian will affect the freezing process, which is rooted in the presence (and the stability) of a long region having  $\ell_{\rm cluster}\approx \frac{\ln L}{\ln 2}$ occupied sites. Let us first define the following frozen state in the Anderson basis
\be
|\Psi_{\rm frozen}^{\rm AL}\rangle=| 
    {\cdots0\,0}\underbrace{\,1\,1\,1\,\cdots \,1\,1\,1}_{\ell_{\rm cluster}}{0\,0\cdots} \rangle,
    \label{eq:Psi}
\ee
for which the real-space occupation of the original fermions~$\langle n_j\rangle=\langle c^{\dagger}_{j}c^{\phantom\dagger}_{j}\rangle$ is maximal in the middle of the central region of length $\ell_{\rm cluster}$. We want to study the stability of $|\Psi_{\rm frozen}\rangle$, and ultimately the fate of $\delta_{\rm min}^{\rm typ}$, against off-diagonal perturbing terms in the interacting toy model framework  Eq.~\eqref{eq:simple}.

The dominant perturbation to the frozen state Eq.~\eqref{eq:Psi} is the 3-body contribution, the $2^{\rm nd}$ line of Eq.~\eqref{eq:simple}, having an amplitude $\sim W^{-d}$ at large disorder~\footnote{Here we used the strong disorder expression for $\xi$ which yields $W\sim \exp(1/\xi)$.} for particle exchanges at  $d\ge 1$ the relative distance between creation and annihilation sites. Ignoring the sub-dominant 4-body terms, we conjecture the following perturbative expansion for the frozen state Eq.~\eqref{eq:Psi}\bea
|\Psi_{\rm frozen}^{\rm MBL}\rangle&\sim& a_0{\ket{\ell_{\rm cluster}}}+ a_1{\ket{\ell_{\rm cluster}-1}}\nonumber\\
&+& a_2{\ket{\ell_{\rm cluster}-2}}+\cdots
\label{eq:PsiMBL}
\eea
where ${\ket{\ell_{\rm cluster}-p}}$ corresponds to $|\Psi_{\rm frozen}^{\rm AL}\rangle$ in Eq.~\eqref{eq:Psi} with $\ell_{\rm cluster}-p$ consecutive occupied orbitals.
Here we will not try  to anticipate the breakdown of such an expansion due to resonances, but instead we are interested in the strong disorder limit where Eq.~\eqref{eq:PsiMBL} should remain a good proxy, with  $a_d\sim W^{-d}$.  In this regime, the clustering process is stable, while being slightly reduced by the off-diagonal terms. Hence, we expect a weak perturbative depletion of the cluster sizes, at leading order in $1/W$ following
\be
\frac{\ell_{\rm cluster}^{\rm AL}-\ell_{\rm cluster}^{\rm MBL}}{\ell_{\rm cluster}^{\rm AL}}\sim W^{-2}.
\label{eq:lALMBL}
\ee
The chain breaking process being controlled by the decay 
$\delta_{\rm min}^{\rm typ}\sim \exp\left(-\frac{\ell_{\rm cluster}}{2\xi}\right)$, the relative difference between the MBL freezing exponent $\gamma^{\rm MBL}$ and its non-interacting counterpart $\gamma^{\rm AL}$ should then follow
\be
\Delta_{\gamma}=\frac{\gamma^{\rm AL}-\gamma^{\rm MBL}}{\gamma^{\rm AL}}\sim W^{-2}.
\label{eq:Deltagamma}
\ee
Below, in Sec.~\ref{sec:NUMMBL}, we will test the above expression against numerical simulations.

\begin{figure*}
    \centering
    \phantomsubfloat{\label{fig:DMTyp_MBLXX_a}}
    \phantomsubfloat{\label{fig:DMTyp_MBLXX_b}}
    \phantomsubfloat{\label{fig:DMTyp_MBLXX_c}}
    \phantomsubfloat{\label{fig:DMTyp_MBLXX_d}}
    \phantomsubfloat{\label{fig:DMTyp_MBLXX_e}}
    \includegraphics[width=\textwidth]{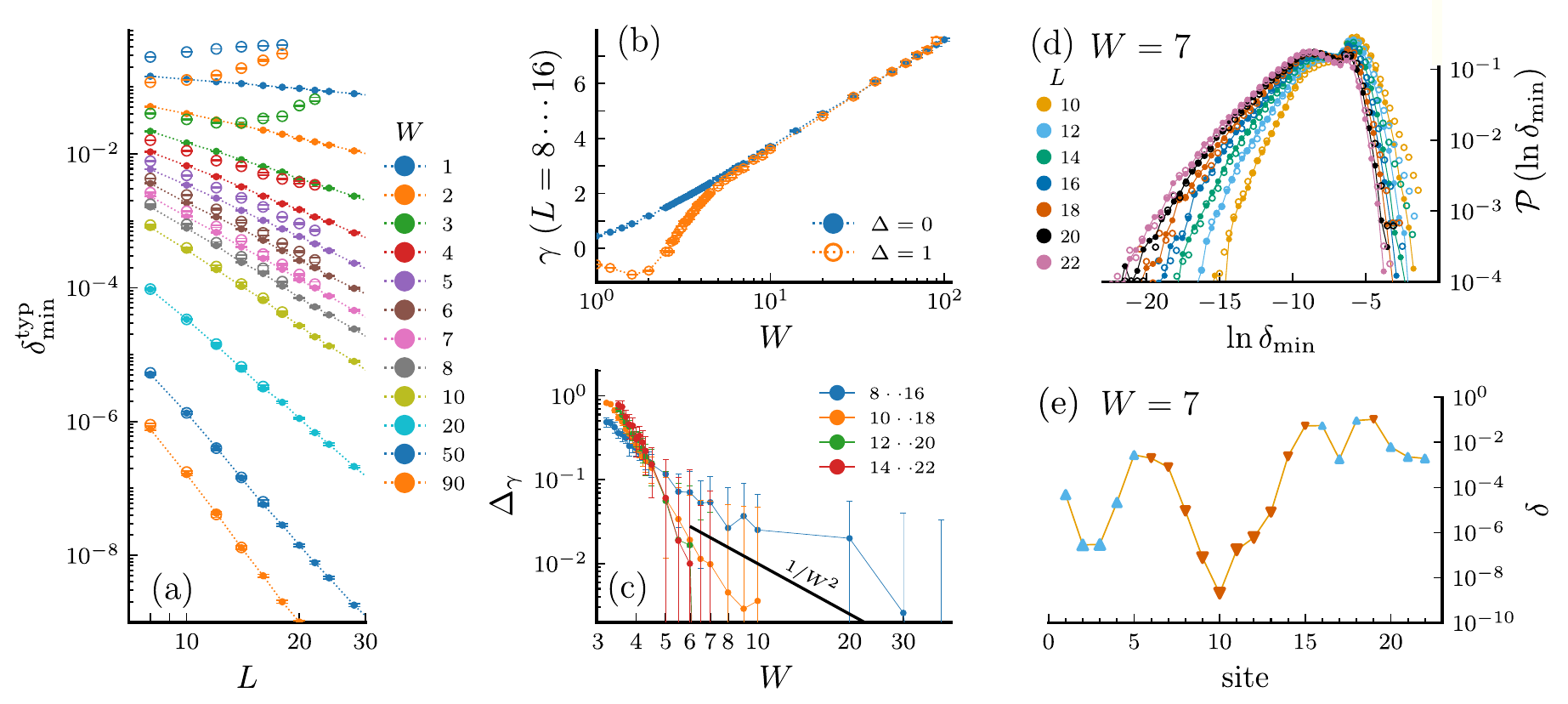}
    \caption{Comparison of the minimal deviations between XX ($\Delta = 0$, full circles, ED results) and Heisenberg models ($\Delta = 1$, open circles, shift-invert ED). \protect\subref*{fig:DMTyp_MBLXX_a}~Finite-size decay of $\delta_{\min}^{\typ}(L)$ plotted for $1\le W\le 90$. The almost perfect agreement at strong disorder between $\Delta=0$ and $\Delta=1$ data slowly deteriorates as the disorder diminishes. \protect\subref*{fig:DMTyp_MBLXX_b}~Disorder-dependence of the freezing exponents $\gamma$ extracted for both cases from finite-window fits to the form  $\delta_{\min}^{\typ} \propto L^{-\gamma}$ with $L=[8\cdots 16]$. The excellent agreement at strong $W$ starts to deviate around $W\sim 5$.
    \protect\subref*{fig:DMTyp_MBLXX_c}~Relative difference between the freezing exponents $\Delta_{\gamma}$ Eq.~\eqref{eq:Deltagamma}, plotted against $W$ for various fitting windows (indicated on the plot). The analytically conjectured strong disorder expansion Eq.~\eqref{eq:Deltagamma} is also show (straight line). \protect\subref*{fig:DMTyp_MBLXX_d}~For $W=7$, comparison of the distributions ${\cal{P}}\left(\ln\delta_{\min}\right)$ for both models (Heisenberg with open symbols {\it{vs.}} XX with lines and solid circles). The agreement improves with increasing system sizes (see also Fig.~\ref{fig:KL} for a quantitative comparison). \protect\subref*{fig:DMTyp_MBLXX_e}~Illustration of the microscopic chain breaking mechanism in the interacting case. Shift-invert ED data for the deviations $\delta_i$ in a $L=22$-site Heisenberg chain. Down (up) triangles denote negative (positive) magnetizations:  a V-shaped cluster is clearly visible, with the most polarized site roughly located in its center.}
    \label{fig:DMTyp_MBLXX}
\end{figure*}

\subsection{Numerical results for minimal deviations and the chain breaking mechanism}
\label{sec:NUMMBL}
In Fig.~\ref{fig:DMTyp_MBLXX} we show shift-invert ED~\cite{luitz_many-body_2015,pietracaprina_shift-invert_2018} results at infinite temperature for the interacting model governed by the random-field Heisenberg Hamiltonian Eq.~\eqref{eq:H}. Our goal is to provide a quantitative comparison with the non-interacting case. Before doing so, we first discuss the microscopic mechanism at play in the interacting model. This is illustrated in Fig.~\ref{fig:DMTyp_MBLXX_e} where the deviations $\delta_i$ are plotted for a single sample ($L=22$ sites, $W=7$). One clearly observes a cluster with a V-shaped form, strongly reminiscent of the non-interacting physics shown in Fig.~\ref{fig:toy1_c}, with the most polarized site been located roughly in the middle of this cluster. Despite the relatively small system sizes available for ED in the presence of interactions, it is reasonable to assume that we are facing the same mechanism as in the free-fermion case.

A more quantitative comparison is provided in Fig.~\ref{fig:DMTyp_MBLXX_a} where the typical deviations are shown for the non-interacting XX (a selection of our ED data already shown in Fig.~\ref{fig:SizeDepDeltaMin}) together with the interacting Heisenberg model results, in both cases for a broad range of disorder strengths $1\le W\le 90$. As already discussed and analyzed in Ref.~\onlinecite{Laflorencie_2020} for Heisenberg, there is a qualitative change in the finite-size scaling  of $\delta_{\rm min}^{\rm typ}(L)$, going from a power-law decay at strong $W$ signalling localization, to an ergodic regime at weaker disorder where  $\delta_{\rm min}^{\rm typ}\to 1/2$. This strong qualitative difference is a smoking gun for the MBL transition, which we believe to be well captured by the extreme polarizations. In Ref.~\onlinecite{Laflorencie_2020} it was further shown that the MBL transition found at $W_c=4.2(5)$ is compatible with a Kosterlitz-Thouless scenario~\cite{PhysRevLett.122.040601,PhysRevB.99.094205,PhysRevB.99.224205,Laflorencie_2020,PhysRevB.102.064207,PhysRevB.106.214202}.

\begin{figure*}[tp]
    \centering
    \phantomsubfloat{\label{fig:DMDist_MBLXX_a}}
    \phantomsubfloat{\label{fig:DMDist_MBLXX_b}}
    \phantomsubfloat{\label{fig:DMDist_MBLXX_c}}
    \phantomsubfloat{\label{fig:DMDist_MBLXX_d}}
    \phantomsubfloat{\label{fig:DMDist_MBLXX_e}}
    \phantomsubfloat{\label{fig:DMDist_MBLXX_f}}
    \includegraphics[width=.95\textwidth]{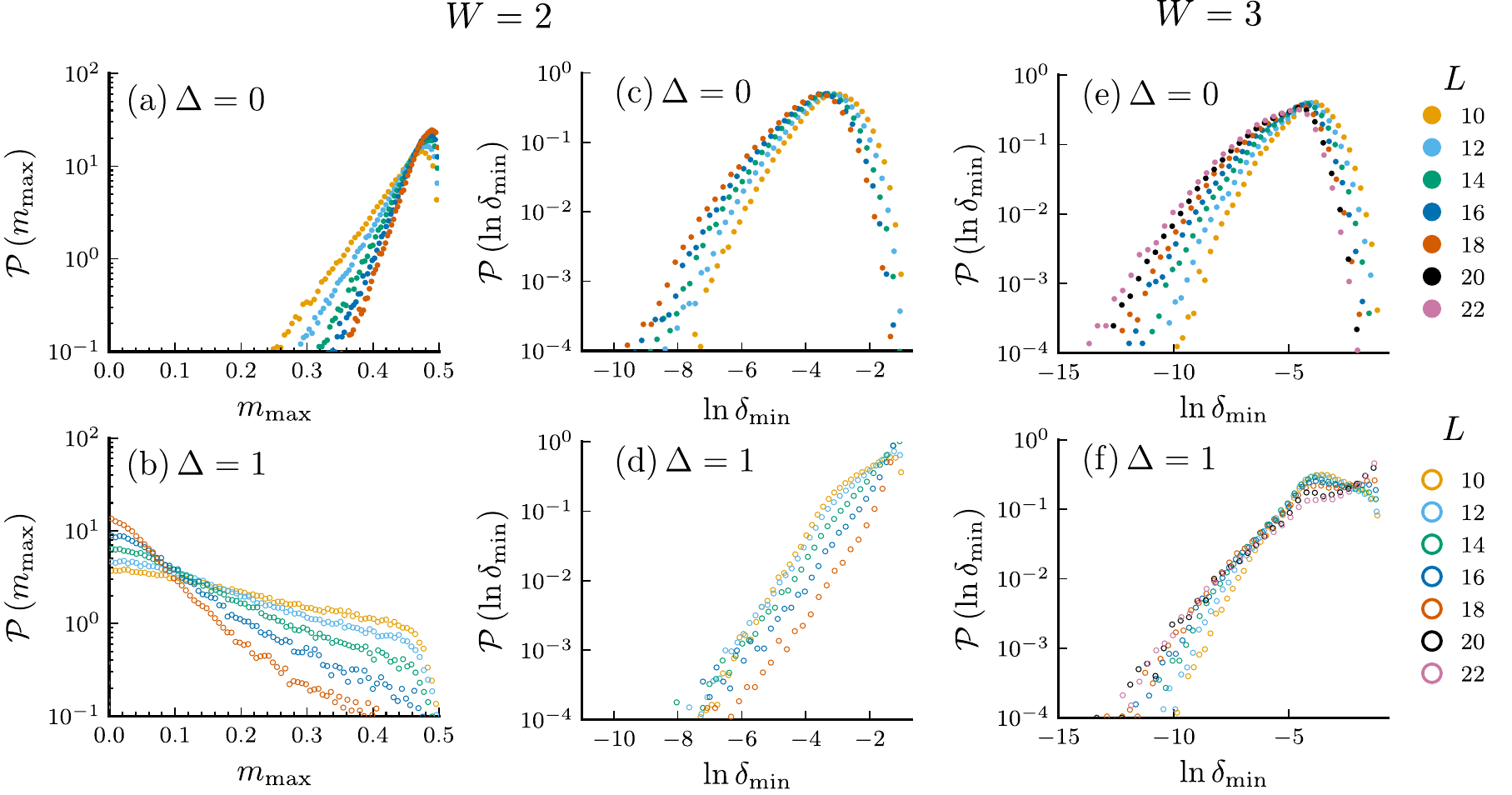}
    \caption{Distribution of the maximal polarization \protect\subref*{fig:DMDist_MBLXX_a},\protect\subref*{fig:DMDist_MBLXX_b}~and the minimal deviations \protect\subref*{fig:DMDist_MBLXX_c}-\protect\subref*{fig:DMDist_MBLXX_f} in the XX chain ($\Delta = 0$, full circles) and in the Heisenberg chain ($\Delta = 1$, open circles). \protect\subref*{fig:DMDist_MBLXX_a}-\protect\subref*{fig:DMDist_MBLXX_d}~At weak disorder, finite size has opposite effects on the distributions of the XX chain~\protect\subref*{fig:DMDist_MBLXX_a},\protect\subref*{fig:DMDist_MBLXX_c} and of the Heisenberg chain~\protect\subref*{fig:DMDist_MBLXX_b},\protect\subref*{fig:DMDist_MBLXX_d}. \protect\subref*{fig:DMDist_MBLXX_e}, \protect\subref*{fig:DMDist_MBLXX_f}~For $W=3$, the distributions have similar exponential tails, but they strongly differ in the large deviations regime.}
    \label{fig:DMDist_MBLvsXX}
\end{figure*}

In Fig.~\ref{fig:DMTyp_MBLXX_a}, the comparison between  Heisenberg (open symbols) {\it{vs.}} XX (closed) is very instructive. At strong disorder, as previously conjectured using  perturbative arguments, the freezing mechanism of the Anderson insulator is only perturbatively ($\sim W^{-2}$) affected by interactions. It is therefore not surprising to observe an almost perfect match of $\delta_{\rm min}^{\rm typ}(L)$ data for the two models at large enough disorder, which appears to be the case for $W\ge 10$. We note however that the non-interacting freezing mechanism is systematically (albeit weakly) altered by interactions:  our  data always obey
\be
\delta_{\rm min~(Heisenberg)}^{\rm typ}- \delta_{\rm min~(XX)}^{\rm typ} \ge 0,
\ee
in agreement with the analytical arguments. When $W$ decreases this difference slowly grows and starts to become qualitative (at the level of our available system sizes) for $W\sim 5$. 

In order to probe this difference in a more quantitative and systematic way, we extract the effective freezing exponents $\gamma$ by fitting  numerical data to the power-law
\be
\delta_{\rm min}^{\rm typ}(L)\propto L^{-\gamma}.
\ee
The exponents $\gamma$ are estimated for both models at $\Delta=0$ (XX) and $\Delta=1$ (Heisenberg) over 5-point fitting windows: $L=[8\cdots 16]~;~[10\cdots 18]~;~[12\cdots 20]~;~[14\cdots 22]$. Results are reported in Fig.~\ref{fig:DMTyp_MBLXX_b} for the first window (for which we have data in the full range of disorder), and for all windows in Fig.~\ref{fig:DMTyp_MBLXX_c} where $\Delta_{\gamma}$, defined  in Eq.~\eqref{eq:Deltagamma}, is plotted against $W$.
First in Fig.~\ref{fig:DMTyp_MBLXX_b} we find a clearly good agreement for $\gamma$ at $\Delta=0{\rm{~and~}}1$ above $W\approx 10$, but below $W\approx 5$ we observe a substantial drop, signalling restoration of ergodicity~\cite{Laflorencie_2020}. The relative difference $\Delta_\gamma$ in Fig.~\ref{fig:DMTyp_MBLXX_c} displays significant variations, and a well marked crossing of the different curves for $W\approx 5$. As a guide, we also show the strong disorder analytical conjecture $1/W^2$  Eq.~\eqref{eq:Deltagamma}, which appears to be in reasonable agreement with the data, at least for the smaller sizes, despite their quite large error bars in this strong $W$ regime. We also note an additional finite-size decrease of $\Delta_\gamma$.

This analysis of the freezing exponents already allows us to draw some partial conclusions. For sufficiently large disorder, typically larger than $W \sim 10$, it seems practically impossible to visually distinguish  Anderson from MBL. Nevertheless, the relative difference $\Delta_\gamma$ remains always  positive, albeit tiny and quickly suppressed with increasing $W$. This suggests that for large disorder the interactions could be an irrelevant perturbation, at least for spin freezing and chain breaking processes. In contrast, as $W$ is reduced the difference begins to be noticeable, with a fairly strong qualitative effect appearing at $W \sim 5$, roughly corresponding  to the regime where the instability of the MBL phase towards  ergodicity  is observed~\cite{luitz_many-body_2015,doggen_many-body_2018,Laflorencie_2020,chanda_time_2020}.

\subsection{Interacting model and extreme value statistics}\label{sec:EVSMBL}
\subsubsection{Distributions}
We want to test the effect of interactions more precisely across the entire Anderson localized regime. The main idea is to go  beyond the typical behavior $\delta_{\rm min}^{\rm typ}(L)$, and exploit our previous results obtained for the full distributions ${\cal{P}}(\delta_{\rm min}^{\rm typ})$. This will allow to better grasp the deep similarities that exist between MBL and AL, and to capture the ergodic instability when reducing $W$.

A first example is provided in Fig.~\ref{fig:DMTyp_MBLXX_d} for $W=7$, where the comparison of the Heisenberg {\it vs.} XX distributions ${\cal{P}}\left(\ln\delta_{\min}\right)$, plotted  for various chains lengths, shows very similar histograms, with all important features well reproduced. Moreover, an essential observation is that  the agreement between the two data sets improves as the system size increases.
This is in sharp contrast  to the weak disorder regime shown for $W=2,\,3$ in Fig.~\ref{fig:DMDist_MBLvsXX}, where the behavior of interacting and non-interacting models appears to be clearly qualitatively different.  Looking more closely, for example for $W = 3$ on the right hand side of Fig.~\ref{fig:DMDist_MBLvsXX}, we observe for the interacting problem in Fig.~\ref{fig:DMDist_MBLXX_f} some signatures of the Gumbel distribution Eq.~\eqref{eq:udist}, which was previously shown in Fig.~\ref{fig:DMDist_MBLXX_e} to describe the log-minimal deviations in the many-body Anderson insulator, and in particular the exponential tail at large negative values of $\ln \delta_{\rm min}$. However, two phenomena are observed as $L$ increases: this tail no longer shifts to the left and at the same time evolving features develop on the other side of the support, in complete contrast to the previously discussed case $W=7$ where instead the similarity between the curves improves with $L$.

\subsubsection{Kullbach-Leibler divergences}
In order to go beyond a simple qualitative comparison of two histograms, we want to quantify the differences and similarities between the interacting {\it{vs.}} non-interacting distributions. We therefore introduce a measure of the proximity between two  probability distributions ($p$ and $q$), given by the Kullbach-Leibler (KL) divergence 
\cite{kullback1951information} defined for a discrete set by
\be
{\rm{KL}}(p|q)= \sum_i q_i \ln \frac{q_i}{p_i}.
\label{eq:KL}
\ee
%
\begin{figure}[bp]
    \centering
    \includegraphics[width=\columnwidth]{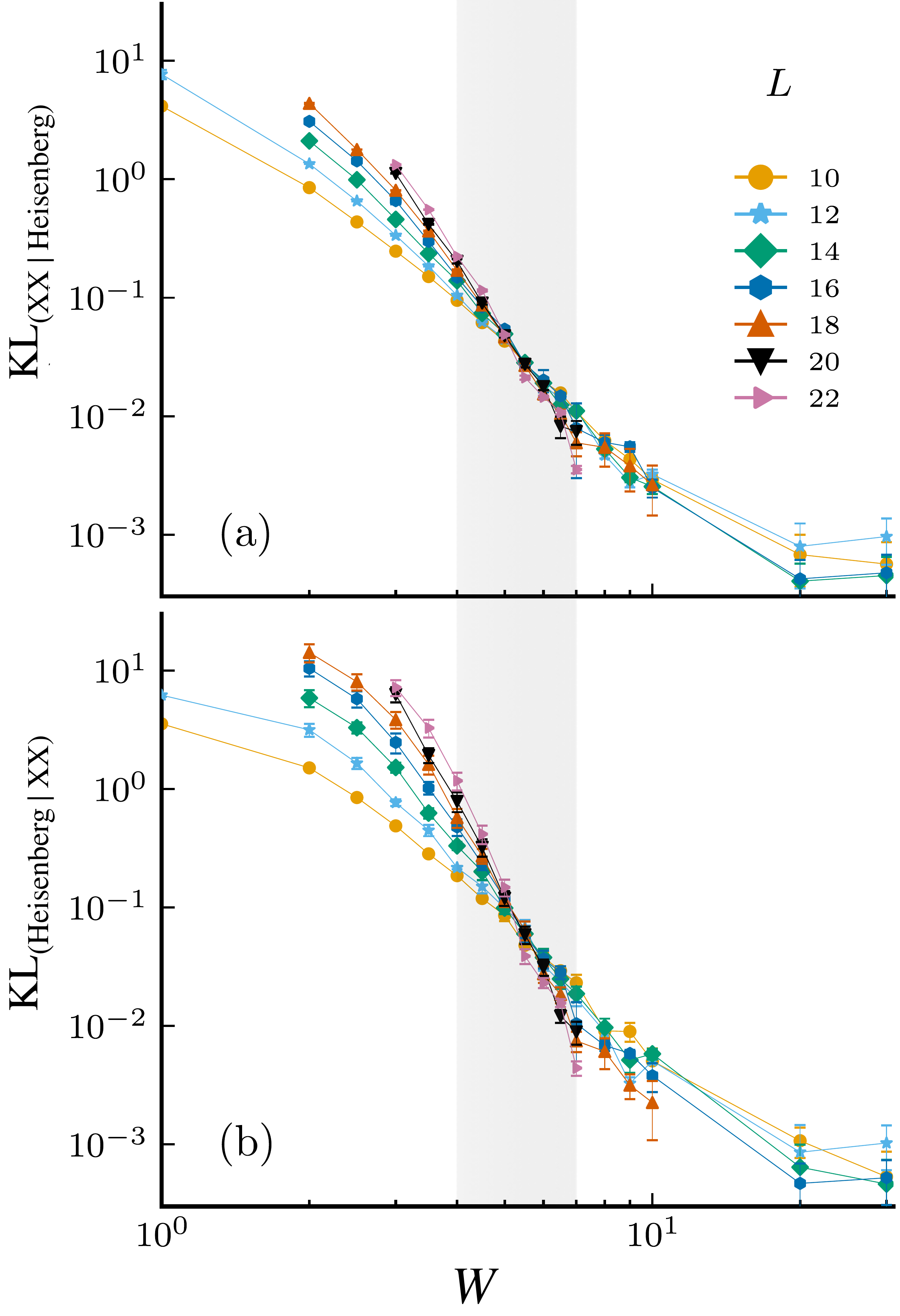}
    \caption{Kullback-Leibler divergences Eq.~\eqref{eq:KL} computed for the distributions of $\ln\delta_{\min}$ for the Heisenberg  {\it{vs.}}  Anderson localized XX chain models. Upon increasing the system length, the distributions become either more and more similar (typically above $W\sim 6-7$), or conversely they  increase their differences (below $W\sim 4$). The two panels show the two (unsymmetrical) versions of KL, both having quantitatively similar behaviors. The grey shaded area corresponds to the extreme-statistics transition regime $W\in [4\,;\,7]$.}
    \label{fig:KL}
\end{figure}
In the limiting cases, KL $\to 0$ if both distributions $p$ and $q$ are similar, while if they differ, KL is non-zero and can be arbitrarily large, possibly diverging (see for instance Refs.~\onlinecite{luitz_many-body_2015,Luitz_2014,PhysRevResearch.2.043346,PhysRevB.106.075107} for a recent use of KL for many-body quantum problems).\\

Therefore, we work with this tool to quantitatively estimate the degree of similarity between the XX and Heisenberg cases, concerning their extreme polarization distributions ${\cal{P}}(\ln \delta_{\rm min})$. Since this object is not strictly speaking a distance measure (not $p\leftrightarrow q$ symmetric) we show the two versions in Fig.~\ref{fig:KL}, where one can easily appreciate the very marked qualitative difference between two regimes. At strong disorder, KL is numerically found to be very small which, as anticipated, signals that MBL and AL yields very similar extreme polarization properties. In contrast, when disorder is reduced we nicely observe a finite-size crossing around $W\sim 4-7$ (gray shaded area in Fig.~\ref{fig:KL}) towards a regime where KL is seemingly diverging with $L$, thus signaling that AL is asymptotically  destabilized by the interactions. 

It is tempting to relate this sharp qualitative change in ${\rm{KL}}_{\rm{(Heisenberg\,|\,XX)}}$ to the MBL-ergodic transition in the Heisenberg model, which has already been extensively discussed in the literature, but never from this perspective.  Indeed it is important to recall that here we are specifically targeting the extreme polarization distributions, and more particularly its comparison with the well-known many-body Anderson localized case realized by the random-field XX chain model. As a matter of fact, such a comparison  provides us a new original  way to address the relevance of interactions, and the conclusion that we can draw here is the following. 
Above, say $W \gtrsim 7$, the extreme statistics of the local spin densities for both Heisenberg and XX are getting extremely close to each other, almost indistinguishable, and this seems to become increasingly true as the system size grows, in agreement with the decrease of $\Delta_{\gamma}$. Our data then strongly suggest that there is an extreme-statistics transition in the regime $W\sim 4-7$. Whether or not this coincides with the MBL transition remains an open question.

\section{Conclusions}
\label{sec:Discussion}
In this work, we have revisited many-particle Anderson and many-body localization physics in one dimension using the new toolbox offered by the extreme value theory applied to this specific problem.
By focusing on the extreme polarizations of two spin chain models in a random magnetic field, the many-body-induced spin freezing mechanism (driving a chain breaking at the thermodynamic limit) has been numerically explored, and compared to an analytically solvable toy model.

In a first part, we have considered the non-interacting limit provided by the XX chain. Using free-fermion exact diagonalization at infinite temperature, the statistics of the local magnetizations, and in particular of their extreme values have been studied in great detail. The emergence of a (disorder-dependent) power-law tail in the distribution of local magnetizations has allowed us to further elaborate on (i) the finite-size scaling of the spin freezing process, which is necessarily accompanied with (ii) a weak link formation leading to chain breaks in the thermodynamic limit. The associated disorder-dependent exponents: spin freezing $\gamma$, weak-link $\theta$, and the  power-law tail  of the distribution $\alpha$, are all simply related, and controlled by the average localization length 
$\xi$. 
Another important result of this non-interacting many-body Anderson limit concerns the distributions of extreme polarizations. Indeed, we found that a single universal distribution is able to describe the entire localized regime, characterized by the Fr\'echet class of generalized extreme value statistics.

Our exact numerics for the multiparticle (half-filled) non-interacting case shows a surprisingly good agreement with the analytically solvable toy model. Firstly, at strong disorder, the power-law decay (with the system size $L$) of the minimal deviations from perfect polarization is perfectly captured by the analysis of the longest sequence of neighboring occupied (or empty) orbitals in real space ($\ell_{\rm max}\sim \ln L$). Nevertheless, this extreme value argument does not exactly describe the entire Anderson localized regime. Indeed, at weak disorder, typically when the localization length $\xi>1$, a new length scale which controls the freezing emerges, namely   
${\overline{\ell_{\rm cluster}}}$ which is the average length of the cluster hosting the most polarized site. Interestingly, an excellent collapse of the minimal deviations data is obtained (without any adjusted parameter) when plotted against ${\overline{\ell_{\rm cluster}}}/\xi$. Finally, the same toy model also provides quantitative results to explain several features in the probability distribution functions.

Lastly, we have explored finite interaction physics and the many-body localization problem from the perspective of extreme value theory. First, based on an interacting toy model, we have conjectured that the non-interacting freezing and chain breaking process should survive interactions in the strong disorder limit. Using state-of-the-art shift-invert diagonalization, we have successfully tested this conjecture (for the available system sizes) by comparing XX and Heisenberg chain data for both the typical extreme deviations and their distributions. More precisely, we then juxtaposed these distributions, and used the Kullbach-Leibler divergence for a quantitative comparison of interacting and non-interacting cases. This allowed us to reveal an extreme-statistics transition in a critical disorder regime $W\sim 4-7$. This could coincide with the MBL transition, but we leave it as an open question for further work.

Our study opens up several other directions. The most obvious and direct one concerns the concrete exploitation of the spin-freezing phenomenon, which paves the way for the controlled elimination of degrees of freedom, such as in RG-type approaches at strong disorder~\cite{fisher_random_1994,fisher_critical_1995,igloi_strong_2005}. For this purpose, it would be interesting to further study not only the most polarized site, but also the density and distance between chain breaks. Here, we have focused on one of the simplest local quantities (the on-site magnetization), which already provides a very rich picture. However, one could think of more sophisticated observables, such as entanglement estimates, more complicated off-diagonal correlators, or more non-local objects such as long-distance correlations. In any case, we strongly believe that the use of extreme statistics theory provides a valuable tool to further investigate the fascinating world of quantum disordered systems.

\section*{Acknowledgments}
It is a great pleasure to thank Nicolas Mac\'e and Gabriel Lemari\'e for  a fruitful initial collaboration on a related work. We also thank  Mari Carmen Ba\~nuls and Lo{\"{\i}}c Herviou for interesting discussions.  This work has been supported by the EUR grant NanoX No. ANR-17-EURE-0009 in the framework of the ''Programme des Investissements d'Avenir''. We acknowledge the use of HPC resources from CALMIP (grants 2022-P0677).

\bibliography{References}

\vskip 1cm
\setcounter{section}{0}
\setcounter{secnumdepth}{3}
\setcounter{figure}{0}
\setcounter{equation}{0}
\renewcommand\thesection{S\arabic{section}}
\renewcommand\thefigure{S\arabic{figure}}
\renewcommand\theequation{S\arabic{equation}}

\newpage
\onecolumngrid
\begin{center}
    \bfseries\Large Supplemental material
\end{center}

Here, we provide some details on more technical aspects of the calculations: the fermionic description in Sec.~\ref{sec:AppFermionic}, the different definitions of the non-interacting localization lengths in Sec.~\ref{sec:AppLL}, the extreme value theory in Sec.~\ref{sec:AppEVT}, and the toy model in Sec.~\ref{sec:AppToyModel}.

\section{Fermionic description of the spin chain model}
\label{sec:AppFermionic}
\subsection{Jordan-Wigner transformation and models}
\label{sec:AppJW}
The Jordan-Wigner transformation~\cite{Jordan_1928} maps spin-$\frac{1}{2}$ operators to Dirac fermions
\bea
n_i&=&\frac{1}{2}+S_i^z\\
c_i^{\vphantom{\dagger}}&=&\exp\left[i\pi \sum_{j=1}^{i-1}n_j\right]S_i^-\\
c_i^{\dagger}&=&S_i^+\exp\left[-i\pi \sum_{j=1}^{i-1}n_j\right].
\eea
Therefore the  spin-$\frac{1}{2}$ XXZ model in a random-field  is equivalent to a chain of interacting spinless fermions with random on-sites energies:
\begin{align}
	\label{eq:HXXZ}
	\mathcal{H}_{\rm XXZ} &= \sum_{i}\Bigl[J (S_i^x S_{i+1}^x + S_i^y S_{i+1}^y + \Delta S_i^z S_{i+1}^z) - h_i S_i^{z}\Bigr]\nonumber\\
    &= \sum_{i} \Bigl[\frac{J}{2} \left(c_i^{\dagger} c_{i+1}^{\vphantom{\dagger}} + c_{i+1}^{\dagger}c_i^{\vphantom{\dagger}}+2 \Delta n_i n_{i+1} \right) -h_i n_i\Bigr]\nonumber\\
    &+ \mathcal{H}_B+\text{C}.
\end{align}
C is an extensive (but irrelevant) constant, and the term $\mathcal{H}_B$ is a boundary contribution coming from the Jordan-Wigner string. It only plays a role for periodic boundary conditions (PBC) and  depends on $N_f$, the number of fermions in the system, such that
\begin{equation}
\label{eq:HB}
	\mathcal{H}_B = \begin{cases}
	-\frac{J}{2}e^{-i\pi N_{f}}(c_1^{\dagger}c_L^{\vphantom{\dagger}} + c_L^{\dagger}c_1^{\vphantom{\dagger}}) &\text{ PBC}\\
	0 &\text{ OBC}
	\end{cases}\quad .
\end{equation}
The Hamiltonians in Eq.~\eqref{eq:HXXZ} have a $U(1)$ symmetry corresponding to the total magnetization being conserved and the total charge being conserved, respectively; throughout this paper, we consider the spin-$\frac{1}{2}$ chain in the $S^{z}_{\mathrm{tot}} = 0$ sector (half-filling in the fermion language).

\subsection{Exact diagonalization for free fermions}
\label{sec:ff}
When $\Delta = 0$, the Hamiltonian in Eq.~\eqref{eq:HXXZ} is a quadratic (free) fermion problem which can be numerically diagonalized for large system sizes. New fermionic operators are built $b_m=\sum_{i=1}^{L}\phi_m(i)c_i$, such that ${\cal{H}}$ is diagonalized:
\be
{\cal{H}}=\sum_{m=1}^{L}{\cal{E}}_m b_m^{\dagger} b_{m}^{\vphantom{\dagger}}.
\ee
All single-particle orbitals $\phi_m(i)$ are spatially localized for any finite disorder strength.
The XX chain can be studied in the $S_{\rm tot}^{z}=0$ sector ($L$ has to be even), which corresponds to $N_f=L/2$ non-interacting fermions. For each disordered sample, the quadratic Hamiltonian is diagonalized, and one fills $L/2$ orbitals $m_1,\,m_2\,\ldots,\, m_{L/2}$. For a given set of occupied orbitals, the total energy is given by
\be
E=\sum_{j=1}^{L/2}{\cal{E}}_{m_j},
\ee
where $\mathcal{E}_{m_j}$ are the single-particle energies. Denoting by $E_{\max}$ and $E_{\min}$ the minimal and maximal energies of the Hamiltonian in the $S^{z}=0$ sector, we can define the energy density above the ground state as
\begin{equation}
	\epsilon = (E-E_{\min})/(E_{\max}-E_{\min}).
\end{equation}

\subsection{Correlations}

Using Wick's theorem to express the correlations efficiently in terms of fermions, we obtain a simple expression in terms of the single-particle wavefunctions of the occupied orbitals $\phi_{m_j}$ (for $i \neq l$):
\begin{align}
 C^{zz}_{i,l}&= \langle S^z_{i}S^z_{l} \rangle -\langle S^z_{i}\rangle\langle S^z_{l} \rangle\\
 &= - \langle c_{i}^{\dagger}c_{l}\rangle \langle c_{l}^{\dagger}c_{i}\rangle\\
 &= - \left|\sum_{j = 1 }^{L/2} \phi_{m_j}^{*}(i)\phi_{m_j}(l) \right|^2,
 \label{eq:CorrFromOccOrb}
\end{align}
where in the last equation, $j$ labels the occupied orbitals.
We see that in the free fermion chain, the (off-diagonal) correlations must always be negative.

In particular, when there is a resonance between the two sites, then $|C^{zz}_{i,l}| \sim 0.25$.  This plays a role in the distribution of the weak-link correlations, which exhibits a bimodal structure, with most samples having close to zero correlations, and a few samples having a resonance and $|C^{zz}_{\cal LR}| \sim 0.25$. Such resonances occur when two conditions are satisfied: (i) the field $h_{\cal L}$ is one of the closest to the field $h_{\cal R}$, and (ii) the field $h_{i^{\star}}$ is very different from both the other fields. The probability of this occurrence is suppressed with $1/L^2$. This has little effect on the decay of the typical weak-link correlations with system sizes, but for sufficiently high sampling rates, it does result in a decay of for the \emph{average} weak-link correlations dominated by $L^{-2}$.

\section{Localization lengths}
\label{sec:AppLL}

Throughout this work, we have used a disorder-dependent localization length, $\xi(W)$, which we plotted in Fig.~1g (main text). We now discuss how this localization length is defined.

\subsection{Motivation}
\par In the (single-particle) Anderson problem, several related notions characterize the localization of orbitals, as reviewed for instance in Refs.~\onlinecite{Kramer_1993, Muller_2016}. 
Perhaps the most intuitive is the inverse participation ratio (IPR)~\cite{Edwards_1972}, characterizing the number of sites that participate in the wavefunction
\begin{equation}
	{\rm P}^{-1} = \frac{\sum_i |\phi_m(i)|^4}{\sum_i |\phi_m(i)|^2}.
\end{equation}
The IPR is numerically easy to compute for a given solution in a given disorder realization; but it characterizes the extent of the wavefunction rather than its exponential decay. For a given wavefunction, the localization length instead captures how fast  the tail of the eigenstate is suppressed:
\begin{equation}
	|\phi_m(i)|^2 \lesssim \exp\left(-\frac{|i-x_{0}^m|}{\xi_m}\right),
\end{equation}
(where $x_0$ could be not on a site). In general, this localization length $\xi_m$ depends on the particular disorder realization and the particular orbital $\phi_m(i)$, which might be very asymmetric or have resonances.  At the same time, to compute the localization length from the IPR, one has to assume a certain shape for the localized orbitals and invert the equation to extract $\xi$~\cite{Laflorencie_2022}; this construction is sensitive to resonances and, furthermore, the IPR has non-analyticities at the band edges~\cite{Johri_2012}.
Instead, one can use a different proxy to compute numerically the localization length $\xi(E,W)$: the Lyapunov exponent~\cite{Kramer_1993,Crisanti_1993,Slevin_2004,Comtet_2013, Johri_2012}, which we discuss shortly in the next section.

\subsection{Lyapunov exponent and random matrices}
Here we give only a rough discussion of the Lyapunov exponent in 1D disordered systems, as it has been extensively discussed elsewhere: Ref.~\onlinecite{Muller_2016} provides simple exercises, Refs.~\onlinecite{Luck_1992,Beenakker_1997,Comtet_2013} give a general discussion, and for a practical approach for finite-size systems one can look at (e.g.) Ref.~\onlinecite{Slevin_2004}. The essential idea is to use the transfer matrix formulation of the Anderson localization problem in 1D. Indeed, considering a solution $\phi$ of the single-particle Anderson problem, we can relate its amplitude on each site by
\begin{equation}
	\begin{pmatrix}
		\phi^{(l+1)}\\
		\phi^{(l)} 
	\end{pmatrix}
	 = 
	 T_l(\mathcal{E})
	\begin{pmatrix}
		\phi^{(l)}\\
		\phi^{(l-1)}
	\end{pmatrix} 
 	= \prod_{j=1}^l T_j(\mathcal{E})
	\begin{pmatrix}
		\phi^{(1)}\\
		\phi^{(0)}
	\end{pmatrix} 
\end{equation}
where $l$ and $j$ label the sites, and where the transfer matrix from $j$ to $j+1$ for a single-particle energy $\mathcal{E}$ is given by
\begin{equation}
T_j(\mathcal{E}) = 
	\begin{pmatrix}
		\frac{\mathcal{E} +h_j}{J/2}  &-1\\
		1 & 0
	\end{pmatrix}.
\end{equation}
In the remainder of this discussion, we omit the energy $\mathcal{E}$ in the notation of the transfer matrix for simplicity. We immediately get that a solution to the fixed-boundary Schrödinger equation is given through a product of uncorrelated, non-commuting $2\times2$ random matrices with unit determinant. 
From Oseledets' and Furstenberg's theorems~\cite{Crisanti_1993,Kramer_1993, Oseledets_1968, Furstenberg_1963}, defining
\begin{equation}
	M_l = \prod_{j = 1}^l T_j
\end{equation}
there exists a limiting matrix 
\begin{equation}
	\Gamma = \lim_{l\rightarrow \infty} \left(M_l^{\dagger} M_l\right)^{\frac{1}{2l}}
\end{equation}
which has real positive eigenvalues $e^{+\lambda}, e^{-\lambda}$; $\lambda > 0 $ is the (maximum) \emph{Lyapunov exponent}. An alternative definition ~\footnote{For a discussion of the differences and relations between the two definitions, see Appendix A in Slevin \emph{et al.}~\cite{Slevin_2004}; a relation to the transport properties is discussed e.g. in Ref.~\onlinecite{Beenakker_1997}. } is
\begin{equation}
	\label{eq:Lyapunov}
	\lambda := \lim_{l\rightarrow \infty} \frac{\ln(|M_l|)}{l}
\end{equation}
(where $| \cdot|$ denotes the matrix norm), which is more directly related to the self-averaging quantity
\begin{equation}
	\lambda(\mathcal{E}) := \lim_{l\rightarrow \infty} \frac{\ln|\phi^{(l)}(\mathcal{E})|}{l}.
\end{equation}
The Borland conjecture~\cite{Kramer_1993,Borland_1963}, later made rigorous~\cite{Matsuda_1970, Carmona_1982} relates the exponential increase of one of the solution of the Cauchy problem to the exponential localization of the wavefunction; the intuition is essentially that, following the exponentially increasing solution from the left, we must get a matching exponentially decreasing solution from the right. This directly relates $\lambda(\mathcal{E})$ to $1/(2\xi(\mathcal{E}))$ for a given disorder distribution.
 
\subsection{Results}

\begin{figure}
	 \centering
    \includegraphics[width=0.5\textwidth]{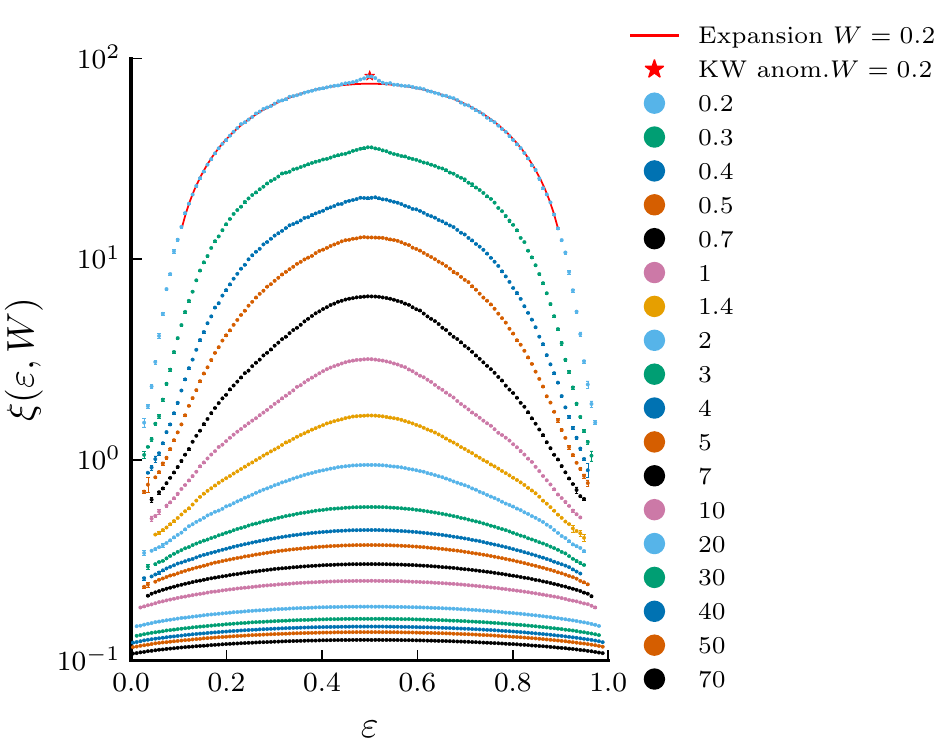}
    \caption{Localization length $\xi(\varepsilon, W)$ from the Lyapunov exponent, where $\varepsilon = (\mathcal{E} - \mathcal{E}_{\min})/(\mathcal{E}_{\max} - \mathcal{E}_{\min})$ is the renormalized single-particle energy.}
    \label{fig:Lyapunov}
\end{figure}

\par In practice, there are several ways of numerically evaluating $\lambda$ in Eq.~\eqref{eq:Lyapunov} (see e.g.~\cite{Muller_2016} for a simple example and~\cite{HernandezMula_2022} for an approach when one has access to extremely large samples). We follow Ref.~\onlinecite{Slevin_2004} in systematically constructing a solution at a given disorder strength and (single-particle) energy $\mathcal{E}$. For each energy, we start with a random initial condition and iterate applying $T_j$ and normalizing until convergence below a  threshold (order of $2 \times 10^{-9}$ for $W = 0.2$ and up to $10^{-5}$ for strong disorder, where less stringent convergence is required). We take a statistical average over 100 samples to have an error estimate (for finite size, there is a distribution of the Lyapunov exponent at given $\mathcal{E}$ and $W$~\cite{Beenakker_1997,Slevin_2004}). To produce Fig.~\ref{fig:Lyapunov}, we sampled $10^3$ energies based on the ED results for a chain of size $L = 2048$ and $10^3$ samples for weak disorder ($W <2$), and $L = 512$ and $10^4$ samples for larger disorder. For readability, we plot the results as a function of the normalized single-particle energies $\varepsilon = (\mathcal{E} - \mathcal{E}_{\min})/(\mathcal{E}_{\max} - \mathcal{E}_{\min})$.

\par The 1D Anderson model admits the following weak-disorder expansion for the localization length as a function of the energy, away from the band edges~\cite{Luck_1992,Kramer_1993,Thouless_1972} (see~\cite{HernandezMula_2022} for a recent summary and numerical verification of those results):
\be
\xi(\mathcal{E}, W)  = \frac{3}{W^2}\left(1-{\cal{E}}^2\right)\quad ({\cal{E}}<1),
\ee
where we used that the variance of the uniform distribution is $V = W^2/3$ and that the jump amplitude is $t = J/2 = 1/2$ in our case. There is also a correction at the band center~\cite{Czycholl_1981} due to the Kappus-Wegner anomaly~\cite{Kappus_1981}:
\be 
\xi(\mathcal{E}=0, W) = \frac{105}{32 W^2}
\ee
Our numerical results fit very well both results at $W = 0.2$, as shown in Fig.~\ref{fig:Lyapunov}.
\par Finally, the results shown in Fig.~1g are obtained by a direct averaging of our numerical results, which by construction amounts to averaging $\xi(\mathcal{E},W)$ over the density of states. We note that the well known behavior at weak disorder $\left(\xi \sim \frac{1}{W^2} \right)$ and at strong disorder $\left(\xi\sim \frac{1}{2\ln(W)}\right)$ are well-captured by the expression~\cite{Potter_2015} $\xi =\left[ \ln\left(1 + \left(\frac{W}{W_0}\right)^2\right)\right]^{-1}$ with $W_0 \in [1.13, 1.22]$ as discussed in Sec.~III (main text).

\section{Extreme value theory}
\label{sec:AppEVT}

In Sec.~IID, we have discussed the extreme value theory for the distribution of the minimal deviations. 
In this section we review some basic results of extreme value theory for applications in the present work.
We refer the interested reader to Ref.~\onlinecite{Majumdar_2020} for a short introduction and to Ref.~\onlinecite{DeHaan_2006} for a detailed mathematical discussion.
\subsection{Limit extreme value distributions}
\subsubsection{Generalized extreme value distribution}
Univariate extreme value theory~\cite{DeHaan_2006} predicts that the law controlling the maximum of $L$ i.i.d random variables $\{x_i\}_{i= 1, \dots, L}$ is controlled in the limit of $L\rightarrow \infty$ by the tail of the parent distribution of these variables. There are three families of limiting distributions which can be gathered in the same \emph{generalized extreme value} (GEV) distribution (Eq.~(2.16) in the main text). The criteria are treated in detail in, for instance, Ref.~\onlinecite{DeHaan_2006}, and useful examples are given also in Ref.~\onlinecite{Majumdar_2020}. Qualitatively, the three cases are
\begin{enumerate}
 	\item If the parent distribution has an exponential tail, the distribution of the maxima belongs to the attraction basin of the Gumbel law. This is what we expect for the maximal magnetizations in the ergodic regime.
    \item If the parent distribution has a power-law dependence with an upper bound for the support (as is the case for the magnetization in the XX chain or in the XXX chain at strong disorder), the Weibull law applies ($s < 0$). This is what is to be expected for the maximal magnetizations in the localized phases. 
   \item If the parent distribution has a (heavy) power-law tail at large values, then the Fr\'echet law applies.
\end{enumerate}

Two forms are of particular interest to us. First, the Weibull case corresponds to an upper-bounded parent distribution with a power-law behavior~\cite{Majumdar_2020} and can be rewritten as:
\be
	f^W_{\beta, \sigma, \mu}(x) = \frac{\beta}{\sigma} \left( \frac{\mu-x}{\sigma} \right)^{\beta-1} e^{-((\mu-x)/\sigma)^\beta}
\ee
Second, the Gumbel distribution can be rewritten as
\be
	f^G_{\sigma, \mu}(x) =  \frac{1}{\sigma} \exp\left[- \frac{x -\mu}{\sigma} - e^{- \frac{x - \mu}{\sigma}}\right].
\ee

In particular, in Sec.~IID, we discussed the fact that $\delta_{\min}$ follows a law related to the Fr\'echet distribution. An immediate consequence is that  $v = -\ln(\delta_{\min})$ follows a generalized Gumbel distribution. Indeed, using $\gamma = (\alpha+1)^{-1}$ we get
\begin{align}
\mathcal{P}_L(v) &= AL\exp\left(-v/\gamma - AL\gamma e^{-v/\gamma}\right) \\
&=\frac{1}{\gamma} \exp\left[-\frac{v-\gamma \ln(AL\gamma)}{\gamma} - \exp\left(-\frac{v-\gamma \ln(AL\gamma)}{\gamma}\right) \right],
\end{align}
and we can identify the Gumbel distribution with scale $\sigma = \gamma$ and location $\mu = \gamma \ln(AL\gamma)$, as stated in Eq.~(2.24).

\subsubsection{Minimal deviations from a Weibull distribution for the maximal magnetization}
In the main text we have sketched a short derivation of the law for the minimal deviation, related to the Fr\'echet law (in the thermodynamic limit for iid variables). An alternative derivation can be done by realizing that the maximal deviation $m_{\max} = \max_{i} m_i$ must satisfy a (reverse) Weibull distribution with an upper limit at $m_{\max} = 1/2$; we can approximate this constraint on the GEV as
\begin{equation}
    \label{eq:constmu}
    1/2 = \mu - \sigma/s, \quad s < 0.
\end{equation}
Identifying
\begin{equation}
    \begin{cases}
        s &= - 1/(\alpha+1) < 0\\
        y &= - (1 + s x) = \frac{1}{\sigma(\alpha+1)} (m_{\max}-1/2)< 0
    \end{cases}
\end{equation}
gives the expression of the distribution of the $m_{\max}$ as a reverse-Weibull distribution $f^W_{\alpha+1, \sigma(\alpha+1), 1/2}(m_{\max}) dm_{\max}$. Identifying 
\begin{equation}
    \begin{cases}
        \delta_{\min} &= 1/2 - m_{\max}\\
        AL &= (\alpha+1) (\sigma (\alpha+1))^{-(\alpha+1)}
    \end{cases}
\end{equation}
yields
\begin{align}
    \mathcal{P}_L(\delta_{\min}) d \delta_{\min} &= f^W_{\alpha+1, \lambda, 1/2}(m_{\max})d m_{\max}\\
    &= AL\delta_{\min}^{\alpha} \exp\left(-\frac{AL}{\alpha+1} \delta_{\min}^{\alpha+1}\right) d\delta_{min}
\end{align}
recovering the expected result. In the end, the distribution for $\delta_{\min}$ is given by the GEV distribution for $m_{\max} = 1/2 - \delta_{\min}$ 
\begin{equation}
    f_{A,L,\gamma}(\delta_{\min}) d\delta_{\min} = \frac{1}{\sigma} f_{s} \left(\frac{m_{\max} -\mu}{\sigma}\right) dm_{\max}
\end{equation}
with parameters
\begin{equation}
    \begin{cases}
        s &= -1/(\alpha+1) = -\gamma\\
        \sigma &= \gamma(AL\gamma)^{-\gamma}\\
        \mu &= 1/2- \sigma/\gamma = 1/2 - (AL\gamma)^{-\gamma}\
    \end{cases}.
\end{equation}

\subsection{Applicability of EVT}
\begin{figure}
	\centering
	\includegraphics[width=0.45\columnwidth]{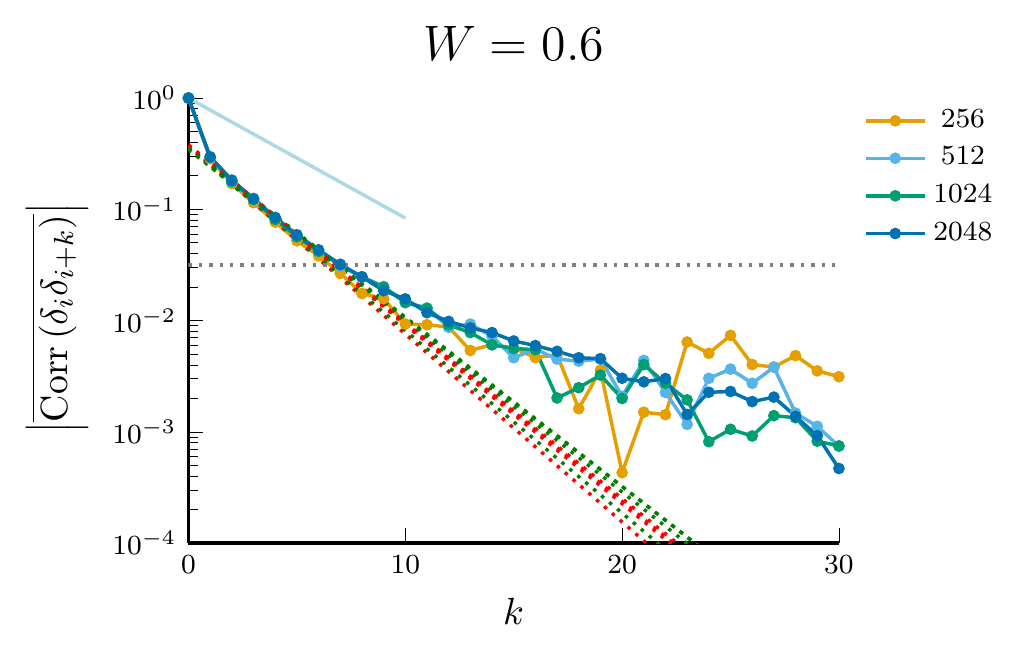}
	\includegraphics[width=0.45\columnwidth]{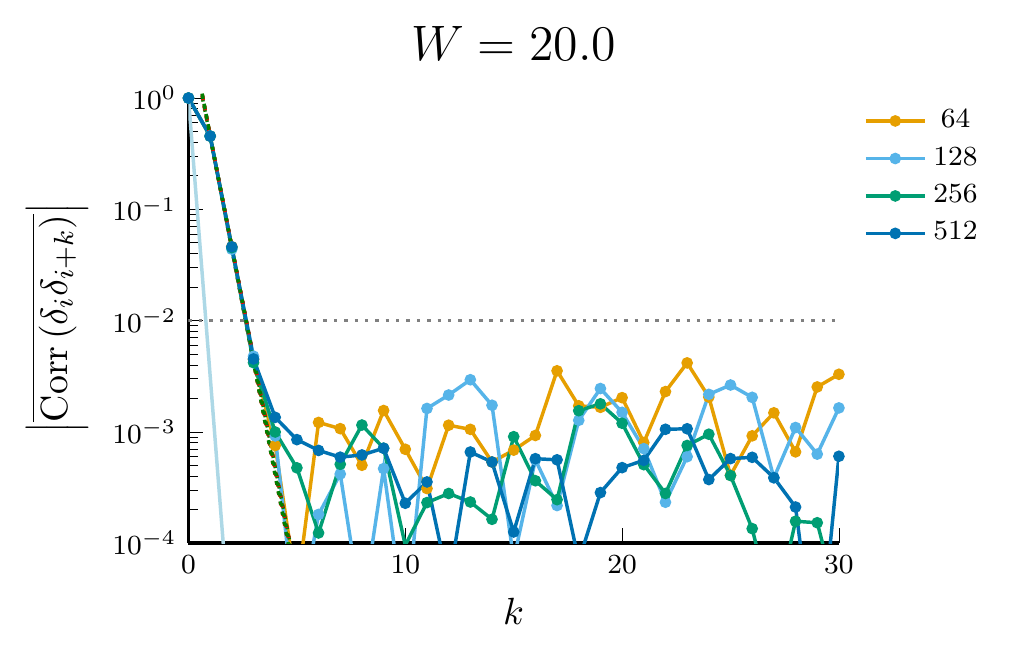}
	\caption{Decay of the correlations between the deviations, for two disorder strengths. The horizontal dotted line is a guide to estimate the threshold below which the values are within errors. The light blue line sketches the decay that would correspond to the localization length. \label{fig:CorrelDelta}}
\end{figure}
In Sec.~IID we have compared our data to the limiting distribution coming from EVT. Here, we show numerical evidence that the correlations between the deviations are indeed weak, decaying exponentially, such that EVT applies to our data. Fig.~\ref{fig:CorrelDelta} shows the correlations between the deviations as a function of the distance $k$, averaged over samples. We note that these correlations decay exponentially. The correlation length is of a similar order of magnitude as the localization length, illustrated by the light blue line in that figure, but it is smaller at weak disorder and seems slightly larger at strong disorder, although it is difficult to evaluate it precisely.

\subsection{Computing $A$ and $\alpha$ from the EV distributions}
\begin{figure}
\centering
\includegraphics[width = 0.5\columnwidth]{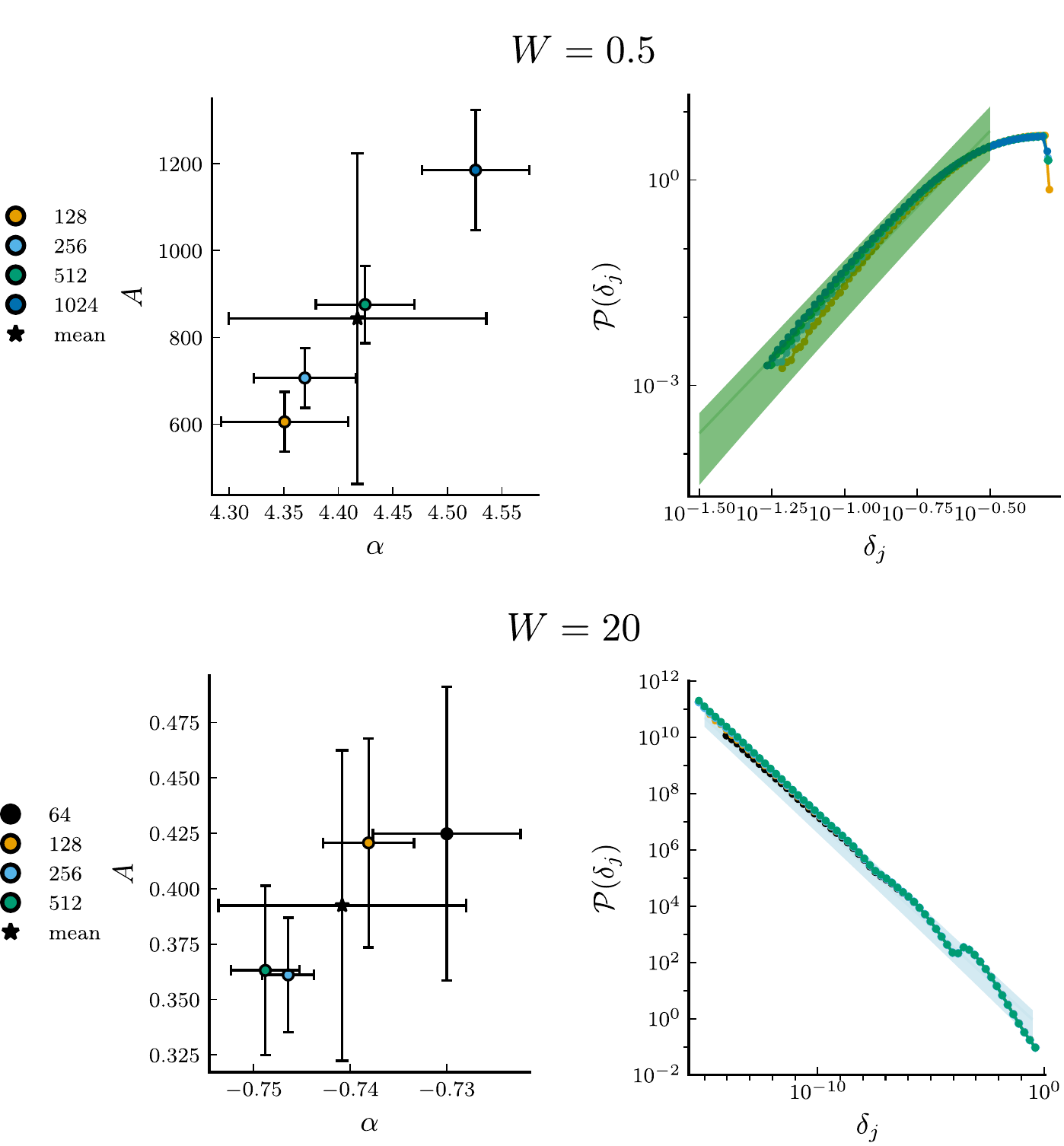}
\caption{\label{fig:AAlpha} Left column : determining $A$ and $\alpha$ from the fits to the extreme value distribution. Right column: comparing the average $A$ and $\alpha$ to the ED results for the distribution of the deviations $\delta$.}
\end{figure}
In Fig.~8 (main text), we have performed fits using the extreme value distribution for $u = \ln(\delta_{\min})$. Further, we have used the values of $A$ and $\alpha$ averaged over the various sizes to collapse the distribution. In Fig.~\ref{fig:AAlpha} we show the values of $A$ and $\alpha$ for the two most challenging cases, at $W = 0.5$ and $W = 20$, where the numerical results do not match perfectly the expected limit distributions. For these two disorders, we show that, although the exponent $\alpha$ has somewhat small fluctuations, the prefactor $A$ has much larger errors. Those result in large corrections in the $\mathcal{P}(\ln \delta)$ distributions; nonetheless, the values of $A$ and $\alpha$ are compatible with the tail of the \emph{full} distribution of the deviations. 
In Fig.~8, the size of the gray area in the collapses is directly determined from the errors shown in Fig.~\ref{fig:AAlpha}. 

\section{Toy model}
\label{sec:AppToyModel}
\subsection{Deviations in the toy model}
In Sec.~III, we discussed the minimal deviations as obtained in the toy model. In this section, we derive some of the key results. 
Eq.~(3.4) (main text) is easily obtained when one considers a cluster of $\ell$ empty (respectively occupied) neighboring orbitals in a sample of otherwise occupied (respectively empty) sites. Indeed, within the toy model framework, the deviation from perfect occupation at a distance $r$ from the center of a cluster is:
\begin{equation}
\label{eq:DeviationToyModel}
    \delta_r^{(\ell)} = \begin{cases}
e^{-\frac{(k-r)}{\xi}}\cdot \frac{(1+e^{-\frac{2r+1}{\xi}})}{1+e^{-\frac{1}{\xi}}} & \text{ if } \ell = 2k,  \\
e^{-\frac{k+1-r}{\xi}}\cdot \frac{(1+e^{-\frac{2r}{\xi}})}{1+e^{-\frac{1}{\xi}}} & \text{ if } \ell = 2k+1\\
\end{cases}
\end{equation}

Let us show in detail the computation of the deviation $\delta_{r}^{(2k)}$ for this case, for a distance $r$ from the center measured as in Fig.~9. Introducing $x = e^{-1/\xi}$ to simplify the expressions, we have
\begin{align}
	\delta_r^{(2k)} &= n_r^{(2k)} = \frac{1-x}{1+x}\left[\sum_{d = k-r}^{k+r} x^d + 2\sum_{d = k+r+1}^{\infty}x^d \right]\\
	&= \frac{1-x}{1+x}\left[ x^{k-r}\frac{1-x^{2r+1}}{1-x} + 2\frac{x^{k+r+1}}{1-x} \right]\\
	&= x^{k-r} \frac{1+x^{2r+1}}{1+x}.
\end{align}
and the calculation for clusters of size $\ell= 2k+1$ yields
\begin{align}
	\delta_r^{(2k+1)} &= \frac{1-x}{1+x}\left[\sum_{d = k+1-r}^{k+r} x^d + 2\sum_{d = k+r+1}^{\infty}x^d \right]\\
	& = \frac{1-x}{1+x} \left[\frac{x^{k+1-r} - x^{k+r+1}}{1-x} +2 \frac{x^{k+r+1}}{1-x}\right]\\
	&= x^{k+1-r}\frac{1+x^{2r}}{1+x}
\end{align}
which directly gives the results in Eq.~\eqref{eq:DeviationToyModel}. The results are the same for an occupied cluster in a sea of empty sites.

As a result, in the center of the cluster we have
\begin{equation}
	\label{eq:MinimalDeviationToyModel}
    \delta_{\min}^{\mathrm{upper}}(\ell) = 
    \begin{cases}
\delta_0^{(2k)}  = e^{-\frac{k}{\xi}}  &\text{ if } \ell = 2k  \\
\delta_0^{(2k+1)} = \frac{e^{-\frac{(2k+1)}{2\xi}}}{\cosh\left(\frac{1}{2\xi}\right)}  &\text{ if } \ell = 2k+1
	\end{cases}
\end{equation} 
corresponding to Eqs.~(3.4) and~(3.5) in the main text (note that in the main text, we consider $\ell = 2k-1$).

The assumption of a cluster of empty (respectively occupied) sites in a sea of sites with the opposite occupation yields an overestimated value for $\delta_{\min}$. If instead we consider the other extreme - a cluster defined by empty sites surrounded by \emph{only} two occupied sites, we obtain
\begin{align}
	\tilde{\delta}_{r}^{(2k)} &=  \frac{1-x}{1+x}\left[x^{k-r} + x^{k+1+r} \right]\\
	\tilde{\delta}_{r}^{(2k+1)} &=  \frac{1-x}{1+x}\left[x^{k+1-r} + x^{k+1+r} \right] 
\end{align}
hence, the minimal deviations are given by
\begin{equation}
	\label{eq:MinimalDev2}
	\delta_{\min}^{\mathrm{lower}}(\ell) = \begin{cases}
	\tilde{\delta}_{0}^{(2k)} &=  x^{k}(1-x),\\
	\tilde{\delta}_{0}^{(2k+1)} &=  2x^{k+1}\frac{1-x}{1+x},
	\end{cases}
\end{equation}
and therefore $\delta_{\min}^{\mathrm{lower}}(\ell) = (1-e^{-\frac{1}{\xi}}) \delta_{\min}^{\mathrm{upper}}(\ell)$, corresponding to Eq.~(3.6).

\subsection{Distributions in the toy model}
\subsubsection{Recovering power-law tails}
With the expressions in Eq.~\eqref{eq:DeviationToyModel} in hands, the power-law tail of the distribution of $\mathcal{P}(\delta)$ in the toy model is easily recovered. Recall that at half-filling, the probability of having a cluster of length $\ell$ is essentially $\mathcal{P}(\ell) \propto 2^{-\ell}$ (there are small corrections but they are not relevant for us here). 
We can roughly estimate the probability of a deviation $\delta$ by 
\begin{equation}
	\mathcal{P}_L(\ln(\delta)) \sim \frac{\xi}{2}\left( \sum_{\{k,r\}'} \mathcal{P}_L(2k)+\sum_{\{k,r\}''} \mathcal{P}_L(2k+1) \right),
\end{equation}
where the prime and second symbols correspond to summing over integers $k,r$ such that
\begin{align}
	\{k,r\}' &= \{k, r| r \leq k \text{ and } \delta_{r}^{2k} \sim \delta\}\\
	\{k,r\}'' &= \{k, r| r \leq k \text{ and } \delta_{r}^{2k+1} \sim \delta\}
\end{align}
respectively. Since $\delta_r^{2k} \sim \delta$ can only happen for $k \geq - \xi \ln(\delta)$, one finds that
\begin{equation}
	\mathcal{P}_L(\ln(\delta)) \propto 2^{-2k} \sim 2^{2\xi\ln(\delta)} = e^{\frac{\ln(\delta)}{\gamma_{\toy}}}
\end{equation}
in agreement with Eq.~(3.3).

\subsubsection{Location of the peaks and merging}
Eqs.~\eqref{eq:MinimalDeviationToyModel} and~\eqref{eq:MinimalDev2} give the distance to be expected between peaks in the distribution of $\ln(\delta_{\min})$ in the toy model.
In particular, one can  directly read off Eqs.~\eqref{eq:MinimalDeviationToyModel} that peaks due to same parity clusters will be spaced by multiples of $1/\xi$.
It is also clear that peaks should occur between isolated odd clusters, and even clusters surrounded by other clusters with a similar occupation. The width of the peaks is thus estimated by
\begin{equation}
	w(\xi) \approx \ln(2) - \ln(1-e^{-\frac{2}{\xi}}).
\end{equation}
With this we can evaluate the order of magnitude of the localization length at which the separated peaks should merge in a single distribution within the toy model: this is when $w \approx 1/\xi$, and gives
\begin{equation}
	\xi = \frac{1}{\ln(1+\sqrt{2})} \approx 1.135,
\end{equation}
which corresponds to a disorder $W \approx 1.4$. This is in agreement with the observations in Fig.~10 (main text).

\subsection{Collapse with $\ln(L)$}
\label{sec:AppToy_collapse}
\begin{figure}
	\centering
	\includegraphics[width = 0.5\columnwidth]{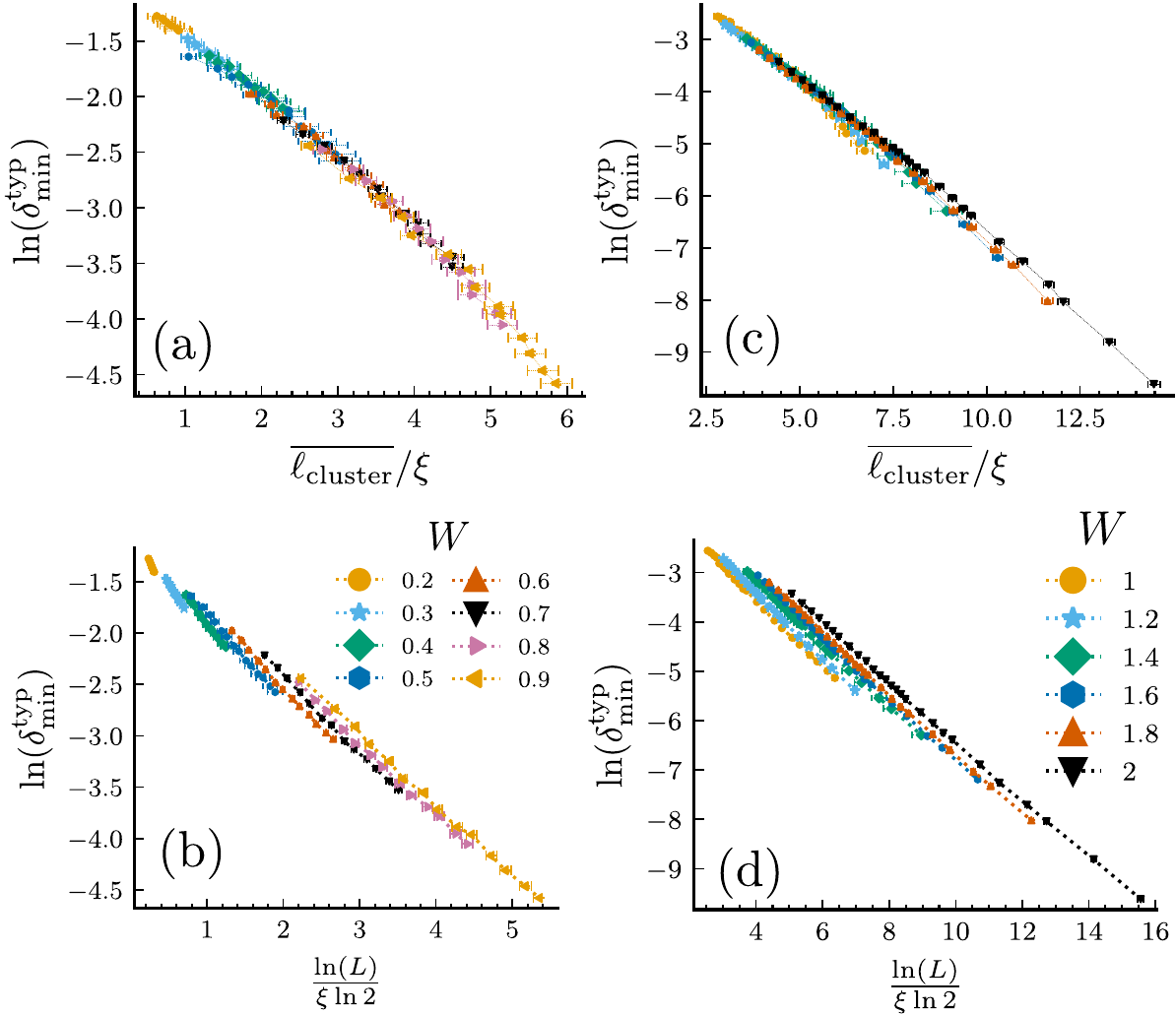}
	\caption{Collapse based on the cluster length $\ell_{\cluster}$ or with $\ln(L)/\ln(2)$. At weak and intermediate disorder, the collapse with  $\ln(L)/\xi$ is not good, unlike the one with $\overline{\ell_{\cluster}}/\xi$ shown in Fig.~9 (main text). \label{fig:CollLogL}}
\end{figure}
In Sec.~III of the main text, we have seen that at weak disorder, the collapse with $\overline{\ell_{\cluster}}$ is surprisingly good, even though this is a limit where the toy model should not apply this well. In Fig.~\ref{fig:CollLogL}, we show that, indeed, the collapse is not as good if we use $\ln(L)/\xi$ instead of $\overline{\ell_{\cluster}}/\xi$, for weak to intermediate disorder strengths.

\end{document}